\documentclass[prb,twocolumn,superscriptaddress,showpacs]{revtex4}

\usepackage{amsmath}
\usepackage{amssymb}
\usepackage{amsfonts}

\usepackage{graphicx}
\usepackage{times}
\usepackage{pstricks}

\newcommand{\be}{\begin{equation} }
\newcommand{\ee}{\end{equation} }
\newcommand{\ba}{\begin{eqnarray} }
\newcommand{\ea}{\end{eqnarray} }

\begin{document}


\title{Majorana Zero Modes in Semiconductor Nanowires in Contact
with Higher-$T_c$ Superconductors}

\author{Younghyun Kim}
\affiliation{Department of Physics, University of California, Santa Barbara, CA 93106}
\author{Jennifer Cano}
\affiliation{Department of Physics, University of California, Santa Barbara, CA 93106}
\author{Chetan Nayak}
\affiliation{Station Q, Microsoft Research, Santa Barbara, CA 93106-6105}
\affiliation{Department of Physics, University of California, Santa Barbara, CA 93106}

\date{\today}

\begin{abstract}
We analyze the prospects for stabilizing Majorana zero modes in
semiconductor nanowires that are proximity-coupled to higher-temperature
superconductors. We begin with the case of iron pnictides which, though they
are $s$-wave superconductors, are believed to have superconducting gaps that
change sign. We then consider the case of cuprate superconductors. We show
that a nanowire on a step-like surface, especially in an
orthorhombic material such as YBCO, can support Majorana zero modes
at an elevated temperature.
\end{abstract}

\maketitle


\section{Introduction}

Majorana zero modes have been predicted to occur in 
certain fractional quantum Hall states
\cite{Moore91,Nayak96c,Lee07,Levin07,Bonderson08a}
and in superconductors in which time-reversal and spin-rotational symmetries
are broken \cite{Volovik99,Read00,Cooper01,Kitaev01,Kitaev06a,Nayak08,Fu08,Sato09,Sato10,Sau10a,Lutchyn10,Oreg10,Alicea11,Fidkowski11}. Very recently, exciting experimental progress
\cite{Mourik12,Rokhinson12,Deng12,Das12}
has been made with semiconductor nanowires \cite{Lutchyn10,Oreg10}
in which superconductivity has been induced by proximity to 
an $s$-wave superconductor (see, in addition, interesting results on a topological
insulator-superconductor interface in Ref. \onlinecite{Williams12}).
The single-particle gap in the nanowire, which is the energy
scale that controls the stability of the zero mode, is determined by
the spin-orbit coupling in the wire, the applied Zeeman field in the direction
of the wire, and the superconducting gap. While the first of these three
quantities is determined primarily by the nanowire and the interface, second
and third are limited primarily by the superconductor. Therefore, it is natural
to consider superconductors with higher-$T_c$s. In this paper, we focus on
the pnictide and cuprate superconductors.

The principal complication with unconventional superconductors
such as the pnictides and cuprates, vis-a-vis the proximity effect,
is that the superconducting gap changes sign from one part of the Brillouin zone
to another. Therefore, it is not even clear that such a superconductor can induce
a non-zero superconducting gap in a nanowire. In addition, the hole-doped
cuprate superconductors, at least, have gapless bulk excitations, which would
hybridize with a Majorana zero mode, thereby giving it a finite lifetime.
However, if these issues could be circumvented, then a higher-$T_c$
superconductor could lead to a Majorana zero mode that is protected by
a larger gap, $\Delta$. This, in turn, can lead to a substantially
smaller splitting $e^{-L\Delta/{v_F}}$ or decay rate $e^{-\Delta/T}$
for zero modes.

In this paper, we show that it is possible for a bulk $s_{\pm}$ superconductor
(such as, according to many theories, a pnictide) to induce a large
superconducting gap in a nanowire. As a result of this gap, a Majorana zero mode
is stabilized at each end of the wire. We further show that a $d$-wave superconductor
can induce a large superconducting gap in a nanowire
that is coupled to it on a step-like terraced surface.
If the superconductor has $d+id$ pairing symmetry, then the
wire would support a Majorana fermion zero mode at each end.
(There is no known superconductor with such a gap symmetry, although
one might expect, on symmetry grounds, that an $id_{xy}$ component would
be induced in a $d_{x^2 - y^2}$ superconductor by the application of a magnetic field.)
If the superconductor has $d_{x^2 - y^2}$ pairing symmetry,
then the Majorana fermion zero mode may decay into the bulk. We
compute this decay rate and analyze the conditions under which the
zero mode is stabilized.

For a recent discussion from a different perspective
of topological superconductivity in semiconductor-cuprate
structures, see Ref. \onlinecite{Takei12}. For interesting recent experimental
results finding a large superconducting gap induced in a topological insulator
by proximity to a cuprate superconductor (as proposed in Ref. \onlinecite{Linder10}),
see Ref. \onlinecite{Zareapour12}.


\section{Basic Setup}
\subsection{Action}
We assume that our system consists of a quasi-1D semiconductor nanowire
of length $L_x$ aligned along the $x$-direction, with width $w$ in the $y$-direction and width
$w_z$ in the $z$-direction. We will assume that ${w_z}\ll w\ll L_x$
and neglect motion in the $z$-direction but allow for the possibility of multiple
sub-bands in the $y$-direction. We will consider two nanowire widths:
$w=50\,\text{nm}$, which is technologically possible (e.g. by e-beam lithography),
as well as $w=100\,\text{nm}$, which is the width in recent experiments
\cite{Mourik12,Rokhinson12,Deng12,Das12} on nanowires coupled
to conventional $s$-wave superconductors.
The nanowire is described by the action $S_{NW}$ while 
the superconductor is described by the action $S_{SC}$; these
are shown below. The nanowire is lying on a (001) surface of the superconductor,
and interactions between them are described by $S_T$.
We will assume that the nanowire is pointing in the (100) direction.
In the case of the cuprates, this will be the optimal arrangement
(as will a nanowire pointing in the (010) direction, which is equivalent).
If the wire points in the (110) direction, then it will be aligned with the
nodes and the induced gap will generically vanish.
The effective action $S = S_{NW} + S_{SC} + S_{T}$ can
be written in the form:
\begin{align}
\label{eqn:starting-action}
S_{NW}=&\frac{2\pi}{\beta}\sum_m\sum_{i}\int dk_x\bar{\Psi}_{m,i,{k_x}}\bigl\{-i\omega_m+\alpha {k_x}\sigma_y
\tau_z \nonumber\\
& +\left(\epsilon_{k_x}+\mbox{$\frac{\hbar^2 (\pi i)^2}{2m^\ast w^2}$}-\mu\right)\tau_z
+V_x\sigma_x \bigr\}\Psi_{m,i,{k_x}}\\
S_{SC}=&\frac{2\pi}{\beta}\sum_m\int_\textbf{q}\bar{c}_{m,\textbf{q}}(-i\omega_m+\xi_q\tau_z+\Delta_q\tau_x)c_{m,\textbf{q}}\\
S_T=&\frac{2\pi}{\beta}\sum_{m,i}\int dk_x\int_{\textbf{q}}{t_i}({k_x},\textbf{q})\bar{\psi}_{m,i,{k_x}}c_{m,\textbf{q}}+h.c.
\end{align}
Here, the index $m$ is a Matsubara index, and $i$ is a sub-band index in the nanowire.
In Eq. (\ref{eqn:starting-action}), we have suppressed the spin and particle-hole
indices on $\Psi$ and $c$. We define $\Psi_{m,i,{k_x}}=(\psi_{m,i,{k_x}\uparrow},\psi_{m,i,{k_x}\downarrow},\bar{\psi}_{-m,i,-{k_x}\downarrow},-\bar{\psi}_{-m,i,-{k_x}\uparrow})^T$,
where $\psi_{m,i,{k_x}\alpha}$ is the annihilation operator for an electron
of Matsubara frequency $\omega_m$, momentum $k_x$, and spin $\alpha$
in the $i^{\rm th}$ sub-band of the nanowire. $c_{m,\textbf{q}}$ is defined analogously as
a spinor in spin and particle-hole space.

The functions ${t_i}({k_x},\textbf{q})$ are the tunneling amplitudes
between the superconductor and the $i^{\rm th}$
sub-band of the semiconductor nanowire. 
We assume for simplicity that the electron wavefunction in the $i^{\rm th}$ sub-band
of the nanowire takes the form $\psi(x,y) = \sqrt{\frac{1}{w}}\sin(\mbox{$\frac{\pi y i}{w}$})\,\psi(x)$.
(The normalization is for later convenience.)
The kinetic energy of an isolated nanowire is $\epsilon_{k_x} \equiv \frac{\hbar^2 k_{x}^2}{2m^\ast}$,
and $m^\ast$, $V_x =g\mu_B B_x/2$, $\mu$, and $\alpha$ are, respectively, the effective mass,
Zeeman splitting, chemical potential, and strength of spin-orbit coupling in the nanowire.
For an InSb nanowire, typical values are $m^\ast = 0.015\,m_e$, $g\approx 50$,
and $\alpha=200\,\text{meV \AA}$. We will assume that the
chemical potential $\mu=$ several meV,
and is tunable with gate voltage. In the absence of superconductivity,
the chemical potential is related to the Fermi
momentum $k_{F,i}$ in the $i^{\rm th}$ sub-band according to
\begin{equation}
\frac{\hbar^2k_{F,i}^2}{2m^\ast} + \frac{\hbar^2 (\pi i)^2}{2m^\ast w^2} - \mu
=\sqrt{V_x^2+\alpha^2k_{F,i}^2}
\label{eqn:kfx}
\end{equation}

In this paper, we consider superconductors with higher $T_c$:
cuprate and pnictide superconductors.
We will take the following form \cite{Norman95} for the band structure in
a cuprate superconductor:
\begin{multline}
\xi_q(eV) = {t_0} + {t_1}(\cos{q_x}a + \cos{q_y}a)/2 +
{t_2} \cos{q_x} \cos{q_y}\\ + {t_3}(\cos 2q_xa+\cos 2q_ya)/2
+{t_5}\cos 2q_xa\cos 2q_ya\\
+ {t_4}(\cos 2q_xa\cos q_ya+\cos q_xa\cos 2q_ya)/2
\label{eqn:xi}
\end{multline}
where ${t_0}=0.1305, {t_1}=-0.5951, {t_2}=0.1636,
{t_3}= -0.0519, {t_4}=-0.1117, {t_5}=0.051$
(all in units of electron volts) and $a\approx 4\, \text{\AA}$.
We will take a superconducting gap of the form
$\Delta_q = {\Delta_0}(\cos{q_x}a - \cos{q_y}a)/2$, and we will take
the representative value ${\Delta_0}=30$ meV.

For iron pnictide superconductors, we consider the BaFe$_2$As$_2$ family.
We focus on the inner hole pocket located at $k_F\sim0.1(\pi/a)$, where again $a\approx 4\, \text{\AA}$, with $s$-wave superconducting gap $\Delta_q=12$meV as in
Ba$_{0.6}$K$_{0.4}$Fe$_2$As$_2$\cite{Ding08}.
We take the following simplified form for the energy dispersion of the
inner hole pocket:
\begin{equation}
\xi_q(eV)=0.5(\cos q_xa+\cos q_ya)-0.95
\end{equation}

\subsection{Tunneling between the Nanowire and the Superconductor}
\label{sec:tunneling}

Using the assumed form $\psi(x,y) = \sqrt{\frac{1}{w}}\sin(\mbox{$\frac{\pi y i}{w}$})\,\psi(x)$
for the electron wavefunction in the $i^{\rm th}$ sub-band in
the nanowire, we can write
${t_i}({k_x},\textbf{q})$ in terms of the position space tunneling amplitude $t(\textbf{r},\textbf{r}')$ as follows:
\begin{multline}
{t_i}(k_x,\textbf{q})=\\
\int \!dx\int^{w}_{0}\!dy\int{d^2}\textbf{r}' \,t(\textbf{r},\textbf{r}')\,e^{i{k_x}x-i{q_x}x'}\,
\frac{e^{-i{q_y}y'}\sin(\mbox{$\frac{\pi y i}{w}$})}{\sqrt{w}}.
\end{multline}
Confinement of the wire in the $y$-direction destroys strict conservation of $y$-momentum.
As a result, we often encounter the following expression, which quantifies this
non-conservation:
\begin{multline}
{g_j}({q_y}) = \int^{w}_{0} dy \frac{e^{-i{q_y}y}\sin(\mbox{$\frac{\pi y j}{w}$})}{\sqrt{w}}\\
= - i^{j+1}\,\frac{e^{-iq_y w/2}}{\sqrt{w}}\biggl[\frac{\sin((q_y w-\pi j )/2)}{q_y - \pi j/w}\,-\\
(-1)^j \frac{\sin((q_y w + \pi j)/2)}{q_y+\pi j/w}\biggr]
\end{multline}
As $w$ is increased, this function becomes more sharply peaked around
${q_y}=\pm \pi j/w$. In nanowires of width $w\approx 100\,\text{nm}$,
momentum non-conservation is small compared to the scale of the Fermi momentum in the superconductor since $k_F^{SC}\sim\pi/a\approx 250\pi/w$.

\begin{figure}[h]
\includegraphics[width=8cm]{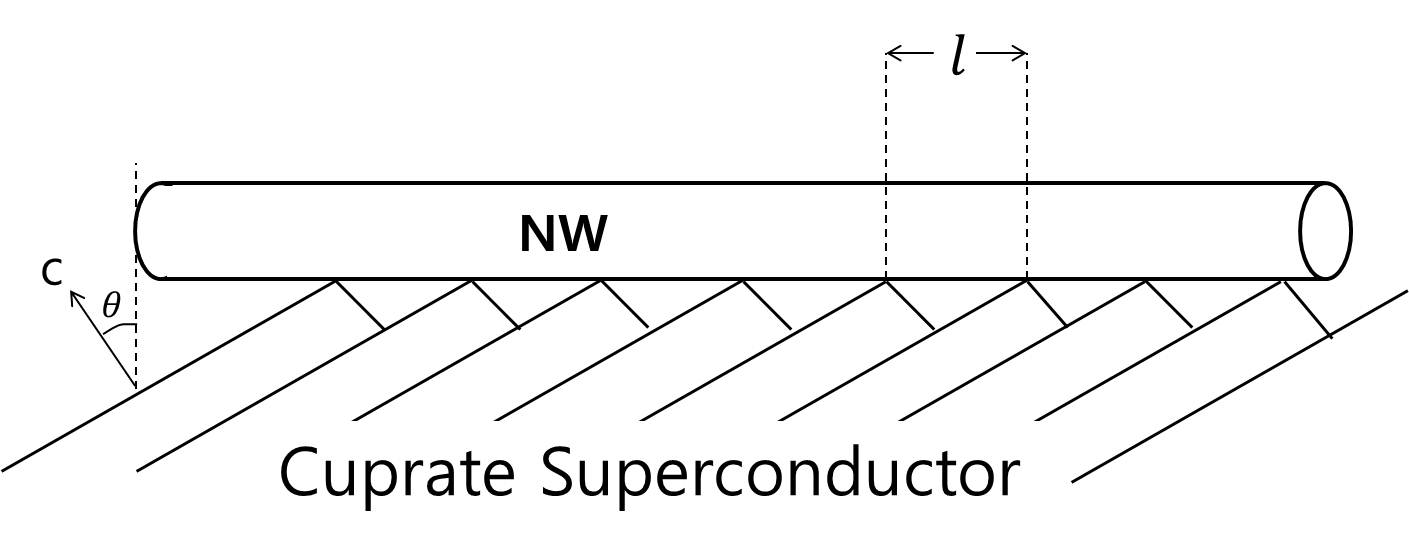}
\caption{Sketch of a interface between a semiconductor nanowire and a step-edge of a cuprate. \label{fig:step-edge}}
\end{figure}

We make further progress by considering several forms of the tunneling matrix elements $t(\textbf{r},\textbf{r}')$
at a clean interface between a higher-$T_c$ superconductor and a semiconductor nanowire,
which we will refer to later:
\begin{enumerate}
\item A smooth, uniform interface (with either a cuprate or pnictide)
where $t(\textbf{r},\textbf{r}')=\frac{t}{2\pi}\delta^{(2)}(\textbf{r}-\textbf{r}')$. Then ${t_i}(k_x,\textbf{q})={t_i^{\text u}}(k_x,\textbf{q})$ with
\begin{equation}
{t_i^{\text u}}(k_x,q)\equiv t\delta(k_x-q_x){g_i}({q_y}).\label{eqn:uniformtunneling}
\end{equation}

\item A smooth uniform interface with a cuprate superconductor such that electrons from the nanowire tunnel only into Cu 4$s$ orbitals. Such a situation is possible because these
are the orbitals that extend furthest in the $z$-directions. Indeed, this is the dominant
path for tunneling between copper-oxide planes in a bi-layer and also through insulating
spacer layers between copper-oxide planes \cite{Andersen95}.
However, depending on the nature of the topmost layers of the cuprate superconductor,
electrons may tunnel instead into other orbitals as well. Therefore, this form of tunneling
presupposes that the topmost (presumably insulating) layers have been
engineered (perhaps through molecular beam epitaxial growth) so that electrons
from the nanowire tunnel through the topmost layers and predominantly into
Cu 4$s$ orbitals in the copper-oxide plane.
The  Cu 4$s$ orbitals hybridize with neighboring Cu 3$d_{x^2-y^2}$ orbitals to
form a (nearly) half-filled band, so that
\begin{multline*}
t(\textbf{r},\textbf{r}') = \frac{t}{2\pi}[\delta^{(2)}(\textbf{r}-\textbf{r}'-a\hat{\textbf{x}})+\delta^{(2)}(\textbf{r}-\textbf{r}'+a\hat{\textbf{x}})\\-\delta^{(2)}(\textbf{r}-\textbf{r}'-a\hat{\textbf{y}})
-\delta^{(2)}(\textbf{r}-\textbf{r}'+a\hat{\textbf{y}})].
\end{multline*}
Then ${t_i}(k_x,\textbf{q}) = {t_i^{\text n}}(k_x,\textbf{q})$ with
\begin{equation}
{t_i^{\text n}}(k_x,\textbf{q}) \equiv 2 t\delta(k_x-q_x){g_i}({q_y})(\text{cos}(q_x a)-\text{cos}(q_ya)).\label{eqn:neighbortunneling}  
\end{equation}

\item A dirty or rough interface. When the interface between the nanowire
and the superconductor is dirty or rough, momentum is not conserved
during the tunneling process.
For illustrative purposes, we consider the extreme case
of $t(\textbf{k},\textbf{q})=\lambda(\textbf{q})$, independent of $\textbf{k}$.
The mismatch between the Fermi momenta of the nanowire and the
superconductor no longer matters, but electrons in the nanowire are now
coupled to different parts of the Brillouin zone, where the gap can have different
signs.

\item A nanowire on top of a step edge of a cuprate, as shown in Figure~\ref{fig:step-edge}.  We assume that the terraces are evenly spaced, with the terrace edges at $x_n=nl$.
We take $l\approx 7$nm, which corresponds to an angle $\theta\approx 10^\circ$.
 In the clean limit, the tunneling amplitude is dominated by the terrace edges $x_n$.
We assume the steps are wide so that tunneling only occurs at $x=x_n$
when $x'<x$; as usual, there is no such restriction for $y$ and $y'$.
Then $t(\textbf{r},\textbf{r}') = \frac{t}{2\pi}\sum_n d\delta(x-x_n)\delta(x\cos\theta-x')\delta(y-y')$, where $d$ is the length scale for the region in which tunneling happens for each step. This yields
${t_i}(k_x,\textbf{q}) = {t_i^{\text s}}(k_x,\textbf{q})$ with
\begin{multline}
{t_i^{\text s}}(k_x,\textbf{q}) \equiv
td\sum_j \delta(k_x-q_x\cos\theta+jQ){g_i}({q_y})
\end{multline}
where $Q=2\pi/l$ and $j$ is summed over the integers.

\end{enumerate}

\subsection{Induced Superconductivity}
\label{sec:induced-superconductivity}

Integrating out the superconductor's degrees of freedom generates
an effective action for the nanowire $S_{\rm eff}=S_{NW}+S'$ where
\begin{align}
S' =&\nonumber\\
-\frac{2\pi}{\beta}&\sum_{m,i,j}\int_{{k_x,k_x'},\textbf{q}}\!\!\bar{\psi}_{m,i,{k_x}}\left[\frac{{t_i}({k_x},\textbf{q})
{t_j}({k'_x},\textbf{q})^{\ast}}{-i\omega_m+\xi_q\tau_z+\Delta_q\tau_x}\right]\psi_{m,j,{k'_x}}\nonumber\\
=&-\sum_{m,i,j}\int_{{k_x,k_x'}}\!\!\bar{\psi}_{m,i,{k_x}}\bigl[ir_{{k_x},{k'_x},i,j,m}\omega_m+\nonumber\\
&\,\,\,\epsilon'_{{k_x},{k'_x},i,j,m}\tau_z+\Delta'_{{k_x},{k'_x},i,j,m}\tau_x\bigr]\psi_{m,j,{k'_x}}
\label{eqn:effective-action}
\end{align}
with
\begin{align}
r_{{k_x},{k'_x},i,j,m} &= \int \frac{{d^2}\textbf{q}}{(2\pi)^2}\frac{{t_i}({k_x},\textbf{q})
{t_j}({k'_x},\textbf{q})^{\ast}}{\omega_m^2+\xi_q^2+
\left|\Delta_q\right|^2},
\label{eqn:wave-renorm}
\\
\epsilon'_{{k_x},{k'_x},i,j,m} &= \int \frac{{d^2}\textbf{q}}{(2\pi)^2}\frac{{t_i}({k_x},\textbf{q})
{t_j}({k'_x},\textbf{q})^{\ast} \, \xi_q}{\omega_m^2+\xi_q^2
+\left|\Delta_q\right|^2},
\label{eqn:energy-renorm}
\\
{\Delta'_{{k_x},{k'_x},i,j,m}} &= \int \frac{{d^2}\textbf{q}}{(2\pi)^2}\frac{{t_i}({k_x},\textbf{q})
{t_j}({k'_x},\textbf{q})^{\ast} \, \Delta_q}{\omega_m^2+\xi_q^2+\left|\Delta_q\right|^2}.
\label{eqn:induced-pairing}
\end{align}
It is important to pause at this point and consider these equations.
They are the same for $s$-wave and $d$-wave superconductors,
but there is a crucial difference in the latter case: the gap $\Delta_q$
may vanish along certain directions in the Brillouin zone and, consequently,
the single-particle energy in the superconductor $\sqrt{\xi_q^2+\left|\Delta_q\right|^2}$
may vanish at certain points. At the nodal points on the Fermi surface of a
$d_{x^2 - y^2}$ superconductor, the denominators in
Eqs. \ref{eqn:wave-renorm}-\ref{eqn:induced-pairing} vanish quadratically
at $\omega_m = 0$. Therefore, these integrals will diverge logarithmically
unless the numerators also vanish. However, for an infinitely-long
nanowire, momentum conservation along the wire (or momentum conservation up to
a multiple of $Q$) prevents any coupling between low-energy electrons and
the nodal points of the superconductor. Consequently, the numerators
in Eqs. \ref{eqn:wave-renorm}-\ref{eqn:induced-pairing} are zero at the
nodal points, where the denominators are dangerous. However, in finite-length
wires, there will be a coupling to the nodal points, and we study this coupling perturbatively
in Section \ref{sec:MZM-decay}.

From the total action $S_{\rm eff}=S_{NW}+S'$, we obtain the
spectrum from the poles of the Green function:
\begin{multline}
\label{eqn:Green-function-eigenvalues}
G^{-1} =
\left(\delta_{ij}\delta_{{k_x}{k'_x}} + r_{{k_x}{k'_x}ij}(\omega) \right) \omega
\\ - \left(\left[\epsilon_{k_x}+\mbox{$\frac{\hbar^2 (\pi i)^2}{2m^\ast w^2}$}-\mu\right]
\delta_{ij}\delta_{{k_x}{k'_x}}
-\epsilon'_{{k_x}{k'_x}ij}(\omega)\right)\tau_z \\
- \alpha {k_x} \sigma_y\tau_z + \Delta'_{{k_x}{k'_x}ij}(\omega)\tau_x
\end{multline}
Here, we have analytically continued $i\omega_m\rightarrow \omega+ i\delta$
and written, e.g. $r_{{k_x}{k'_x}ijm} \rightarrow  r_{{k_x}{k'_x}ij}(\omega+i\delta)$.
The smallest positive pole of this equation is the gap.

If momentum is conserved in the $x$-direction,
then $r, \epsilon', \Delta'$ are all diagonal in $k_x$.
If we further assume the eigenvalues $\omega$ are much smaller than
$\Delta_q$, then we can drop the $\omega$ dependence of $r_{{k_x},i,j}(\omega),
\epsilon'_{{k_x},i,j}(\omega)$, and $\Delta'_{{k_x},i,j}(\omega)$. Finding
the poles of Eq. (\ref{eqn:Green-function-eigenvalues})
reduces to finding the eigenvalues of the matrix $M$:
\begin{multline}
\label{eqn:eigenvalues-simple}
M = \left(\delta_{ij} + r_{ij}({k_x})\right)^{-1}\Bigl(
\alpha {k_x} \sigma_y\tau_z + \Delta'_{ij}({k_x})\tau_x\\
+\,\left(\left[\epsilon_{k_x}+\mbox{$\frac{\hbar^2 (\pi i)^2}{2m^\ast w^2}$}-\mu\right]\delta_{ij}+\epsilon'_{ij}({k_x})\right)\tau_z\Bigr)
\end{multline}
Since $w\gg a$, momentum non-conservation in the $y$-direction
is small on the scale of the Fermi momentum of the superconductor.
We make the following approximation, for the moment, in
order to take a qualitative look at the induced gap (we will use the full expressions
when we turn to a more careful computation of the induced gap):
\begin{equation}
{g_j}({q_y}) \approx
-\frac{\pi i^{j+1} e^{-i{q_y}w/2}}{2i\sqrt{w}} [\delta(q_y-{\pi j/w})-(-1)^j\delta(q_y+{\pi j/w})].
\end{equation}
In this limit, we can simplify
the expressions in the previous subsection for $t^u$ and $t^n$.
Now, the effective action $S_{\rm eff}$ takes the form: 
\begin{multline}
S_{\rm eff}=\frac{2\pi}{\beta}\sum_{m,i}\int dk_x\bar{\Psi}_{m,i,k_x}\bigl[-i(1+r_{k_x,i,m})\omega_m+
\alpha {k_x}\sigma_y \tau_z\\
+(\epsilon_{k_x}+\mbox{$\frac{\hbar^2 (\pi i)^2}{2m^\ast w^2}$}-\mu-\epsilon'_{k_x,i,m})\tau_z\\
+V_x\sigma_x-{\Delta'_{k_x,i,m}}\tau_x\bigr]\Psi_{m,i,k_x}
\label{eqn:effective-action}
\end{multline}
with Eqs. (\ref{eqn:wave-renorm})-(\ref{eqn:induced-pairing}) now taking the simpler form
\begin{align}
r_{k_x,i,m} &= \frac{\left|t_{i,k_x}\right|^2}{\omega_m^2+\xi_{k_x,\pi i/w}^2+\left|\Delta_{k_x,\pi i/w}\right|^2},\\
\epsilon'_{k_x,i,m} &= \frac{\left|t_{i,k_x}\right|^2 \, \xi_{k_x,\pi i/w}}{\omega_m^2+\xi_{k_x,\pi i/w}^2+\left|\Delta_{k_x,\pi i/w}\right|^2},\\
\Delta'_{k_x,i,m} &= \frac{\left|t_{i,k_x}\right|^2 \, \Delta_{k_x,\pi i/w}}{\omega_m^2+\xi_{k_x,\pi i/w}^2+\left|\Delta_{k_x,\pi i/w}\right|^2}.
\label{eqn:induced-pairing2}
\end{align}
We now see that, in the limit in which we replace $g_j(q_y)$ by a sum of $\delta$-functions,
there is no coupling between the nanowire and the nodal points in the superconductor
for generic values of $k_x$ (including the expected Fermi momenta for realistic nanowires). 
In these equations, $t_{i,k_x}$ is given by
\begin{align}
\left|t_{i,k_x}\right|^2 &= \int_{k_x'}\int \frac{d^2\textbf{q}}{(2\pi)^2}t_i(k_x,\textbf{q})t_i(k_x',\textbf{q})^\ast\nonumber\\
&\sim \left|t\right|^2 \mbox{ for } t_i=t_i^u\\
&\sim \left|t\right|^2(\cos k_xa - \cos k_ya)^2 \mbox{ for } t_i=t_i^n
\end{align}

The induced superconducting gap function is:
\begin{align}
{\Delta}^{\rm ind}_{k_x,i,m} &= \frac{{\Delta'_{k_x,i,m}}}{1+r_{k_x,i,m}}\nonumber\\
&=\frac{{\left|t_{i,k_x}\right|^2}{\Delta_{k_x,\pi i/w}}}{\omega_m^2+\xi_{k_x,\pi i/w}^2+\left|\Delta_{k_x,\pi i/w}\right|^2+
{\left|t_{i,k_x}\right|^2}}
\label{eqn:inducedgap}
\end{align}
At the Fermi surface, in the static limit, this is
\begin{equation}
{\Delta}^{\rm ind}_{{k_F},i}  \equiv {\Delta}^{\rm ind}_{{k_F},i,0} 
=\frac{{\left|t_{i,k_x}\right|^2}{\Delta_{k_F,\pi i/w}}}{{\left|t_{i,k_F}\right|^2} + \xi_{k_F,\pi i/w}^2+ \left|\Delta_{k_F,\pi i/w}\right|^2}
\label{eqn:static-inducedgap}
\end{equation}
Note that $\xi_{k_{F,i}}$ may not vanish due to the mismatch between the
Fermi momentum of the nanowire and that of the superconductor.
This mismatch is one of the limiting factors for induced superconductivity.

From the single-particle spectrum obtained from Eq. (\ref{eqn:eigenvalues-simple}),
we see that the single-particle gap at $k_F$ in the $i^{\rm th}$ sub-band is:\cite{Sau12}
\begin{equation}
\Delta^{\rm qp}_{{k_F},i}=\frac{E_{SO}}{\sqrt{V_x^2+E_{SO}^2}}\Delta^{\rm ind}_{{k_F},i}
\label{eqn:qpgap}
\end{equation}
where $E_{SO} = \alpha k_{F,x}$ and $k_{F,x}$ is the Fermi momentum in the
$x$-direction.

\subsection{Renormalized Parameters and Topological Phase Transition}

The effective action $S_{\rm eff}$ takes the form of the action of a nanowire with an applied pair field
$\Delta'_{{k_x},i,m}$, as may be seen in the $w\gg a$ limit in Eq. (\ref{eqn:effective-action}).
However, the parameters $\mu, \alpha, \epsilon, V_x$ are renormalized compared
to those of an isolated nanowire as a result of the coupling to the superconductor.
In the $w\gg a$ limit, this is simply renormalization by a factor of $(1+r)^{-1}$,
as may be seen by making the change of variables
$\bar{\Psi}_{m,i,k_x}\rightarrow \bar{\Psi}_{m,i,k_x} (1+r_{k_x,i,m})^{-1/2}$, $\Psi_{m,i,k_x}\rightarrow \Psi_{m,i,k_x} (1+r_{k_x,i,m})^{-1/2}$ in Eq. (\ref{eqn:effective-action}),
which now takes the form:
\begin{multline}
S_{\rm eff}=\frac{2\pi}{\beta}\sum_{m,i}\int dk_x\bar{\Psi}_{m,i,k_x}\bigl[-i\omega_m+
\alpha_{k_x,i,m}^{\rm ind} {k_x}\sigma_y \tau_z\\
+\left(\epsilon^{\rm ind}_{k_x}-\mu_{k_x,i,m}^{\rm ind}\right)\tau_z
+(V_x)_{k_x,i,m}^{\rm ind}\sigma_x-{\Delta^{\rm ind}_{k_x,i,m}}\tau_x\bigr]\Psi_{m,i,k_x}
\label{eqn:effective-action-rescaled}
\end{multline}
where
\begin{multline}
(V_x)_{k_x,i,m}^{\rm ind} = {V_x}/({1+r_{k_x,i,m}}), \alpha_{k_x,i,m}^{\rm ind} = \alpha/({1+r_{k_x,i,m}}),\\
\mu_{k_x,i,m}^{\rm ind} = (\mu - \mbox{$\frac{\hbar^2 (\pi i)^2}{2m^\ast w^2}$})/({1+r_{k_x,i,m}}).
\end{multline}
In particular, the Zeeman splitting $V_x$ takes a renormalized value.
Since $V_x$ is independent of $|t|^2$ but $r$ is proportional to $|t|^2$,
larger tunneling effectively reduces the $g$-factor of the nanowire. As we will see in subsequent sections, the condition to be in a topological phase sets a lower bound on $V_x^{\rm ind}$.
Hence, a reduced effective $g$-factor would require a larger magnetic field to reach the topological regime. One possible concern is that the larger required magnetic field would destroy superconductivity, but in the high-T$_c$ materials that we consider, the critical field is large enough that this is not the case. Even if we could induce a superconducting gap in the nanowire on the order of the gap in a cuprate superconductor, $\Delta^{\rm ind} \sim $30 meV, the threshold to be in the topological phase would require $V_x^{\rm ind}$ to be at least this large, corresponding to an applied field of about 40T for the values of $r \sim 1$ that we consider in this paper. While this field is very large, it is still less than $H_{c2}$ in a cuprate superconductor. Moreover, as we will see, the induced superconducting gap is typically a few meV and, therefore, easily achievable with
magnetic fields of a few Tesla, which will have negligible effect on a high-$T_c$ superconductor.

When searching for the topological superconducting phase, it is the renormalized parameters and not their counterparts measured before the wire is coupled to the superconductor that determine the onset of the phase. The topological phase is characterized by the existence of Majorana zero modes
at the end of the wire, which occurs when a condition of the following form is satisfied:
\begin{equation}
(V_x)_{k_x=0,i}^{\rm ind} >
\sqrt{\left(\mu_{k_x=0,i}^{\rm ind}\right)^2 +
\left(\Delta_{{k_x=0},i}^{\rm ind}\right)^2}
\end{equation}
The parameters that enter this condition are renormalized.

\subsection{Majorana Zero Modes}

We now consider the possibility of having Majorana zero modes at the end of the wire.
For simplicity, we will assume in this subsection that the Fermi level lies in the
lowest band and take sub-band index $i=1$. We will suppress the sub-band
index and write $\psi \equiv \psi_{i=1}$.
The Hamiltonian of the nanowire can be written in the following form
in real space, where all parameters now correspond to their induced values after coupling to the superconductor, described in the previous section (we drop the superscripts `${\rm ind}$'
in order to avoid clutter):
\begin{multline}
H=\int\!\! dx\Bigl\{ \psi_{\sigma}^{\dagger}(x)\left( -\mbox{$\frac{\hbar^2\partial_x^2}{2m^\ast}$}-\mu(x)+i\alpha\sigma_y\partial_x+V_x\sigma_x\right)\psi_{\sigma'}(x)\\
+\int\!\! dx\,dx'[(\Delta(x,x')\psi^\dagger_\uparrow(x)\psi^\dagger_\downarrow(x')+ h.c.]\Bigr\},
\end{multline}
We assume that the nanowire lies along the negative $x$-axis and terminates
at $x=0$. This condition can be realized by setting $\mu(x<0)=\mu_0$, and $\mu(x\geq0)=-\infty$.

Note that such a BdG Hamiltonian is justified in the $d_{x^2 - y^2}$-wave
case if the nanowire is decoupled from the nodal points of the superconductor.
This is the case in the limit of an infinitely-long wire, as
we saw in Section \ref{sec:induced-superconductivity}.
However, when we consider Majorana zero modes at the ends of
(the topological portion of) a wire, we are necessarily faced with
a situation in which momentum along the wire is not conserved,
so there will be some coupling between the zero modes and
nodal excitations. We set this coupling to zero here and treat it
perturbatively in Section \ref{sec:MZM-decay}.

Then our Hamiltonian in the Nambu basis $\Psi^{\dagger}(x)=(\psi^\dagger_\uparrow(x),\psi^\dagger_\downarrow(x),\psi_\downarrow(x),\psi_\uparrow(x))$ can be written as,
\[
H=\int dx\, \Psi^\dagger(x)H_{BdG}\Psi(x)\]
where,
\begin{multline}
H_{BdG}= \int dx \Bigl[ \delta(x-x')
\left(-\mbox{$\frac{\hbar^2\partial_x^2}{2m^\ast}$}-\mu(x)+V_x\sigma_x\right)\tau_z\\
+ \delta(x-x')i\alpha\sigma_y\partial_x
+ \,\Delta(x,x')\sigma_z\tau_x\Bigr]
\end{multline}
which gives the following BdG equation for $E=0$:
\begin{equation}
H_{BdG}\cdot(u_\uparrow(x), u_\downarrow(x), v_\downarrow(x), v_\uparrow(x) )^T = 0
\end{equation}
Since the BdG Hamiltonian is real, we can have real solutions for Majorana zero modes. After imposing particle-hole symmetry for a real solution, we can set $v_{\uparrow/\downarrow}(x)= \lambda u_{\uparrow/\downarrow}(x)$ with $\lambda=\pm1$. The BdG equation for $E=0$ can be written as,
\begin{multline}\label{eq:bdg}
\int dx' \mbox{$\begin{pmatrix} \scriptstyle{-\delta(x-x')}\left(\mbox{$\frac{\hbar^2\partial_x^2}{2m^\ast}$}+\mu_0\right) & {V_+}(x,x')\\ 	{V_-}(x,x') &
	 \scriptstyle{-\delta(x-x')}\left(\mbox{$\frac{\hbar^2\partial_x^2}{2m^\ast}$}+\mu_0\right) \end{pmatrix}$}\,\times\\
	\begin{pmatrix} u_\uparrow(x') \\ u_\downarrow(x') \end{pmatrix} =0
\end{multline}
where
\begin{equation}
V_{\pm}(x,x')\equiv
V_x  \delta(x-x')\pm\lambda\,\Delta (x,x')\pm\alpha\delta(x-x')\partial_x
\end{equation}
with 3 constraints: $[u_{\uparrow/\downarrow}(x=0)]=0$ and normalization.
Assuming $u_{\uparrow/\downarrow}(x<0)\propto e^{zx}$, the existence of a zero mode
requires at least three roots $z_i$ with positive real part, so that it is normalized
and localized at the end $x=0$.



\section{Superconductors with $s_{\pm}$ Pairing Symmetry}
Recent results on iron pnictide superconductors suggest that its superconducting
order parameter has $s\pm$ pairing symmetry. The Fermi surface has
several components, and the sign and size of the gap
vary from one Fermi surface component to another but the gap
does not change sign as any Fermi surface component is encircled.
The smaller hole pocket centered at the $\Gamma$ point has a gap
$\Delta\approx 12$meV (see, e.g. Ref. \onlinecite{Ding08}).
Therefore, in the clean interface limit with momentum conserving tunneling,
we can induce a large superconducting gap at the Fermi points
of a nanowire by bringing the Fermi momentum of a nanowire near to this
hole pocket's Fermi surface. The Fermi level for the smaller hole pocket near the
$\Gamma$ point ranges from 0.1 to 0.3 $(\pi/a)$ depending on the
doping level \cite{Brouet09}. 
\begin{figure}[h]
\includegraphics[width=9cm]{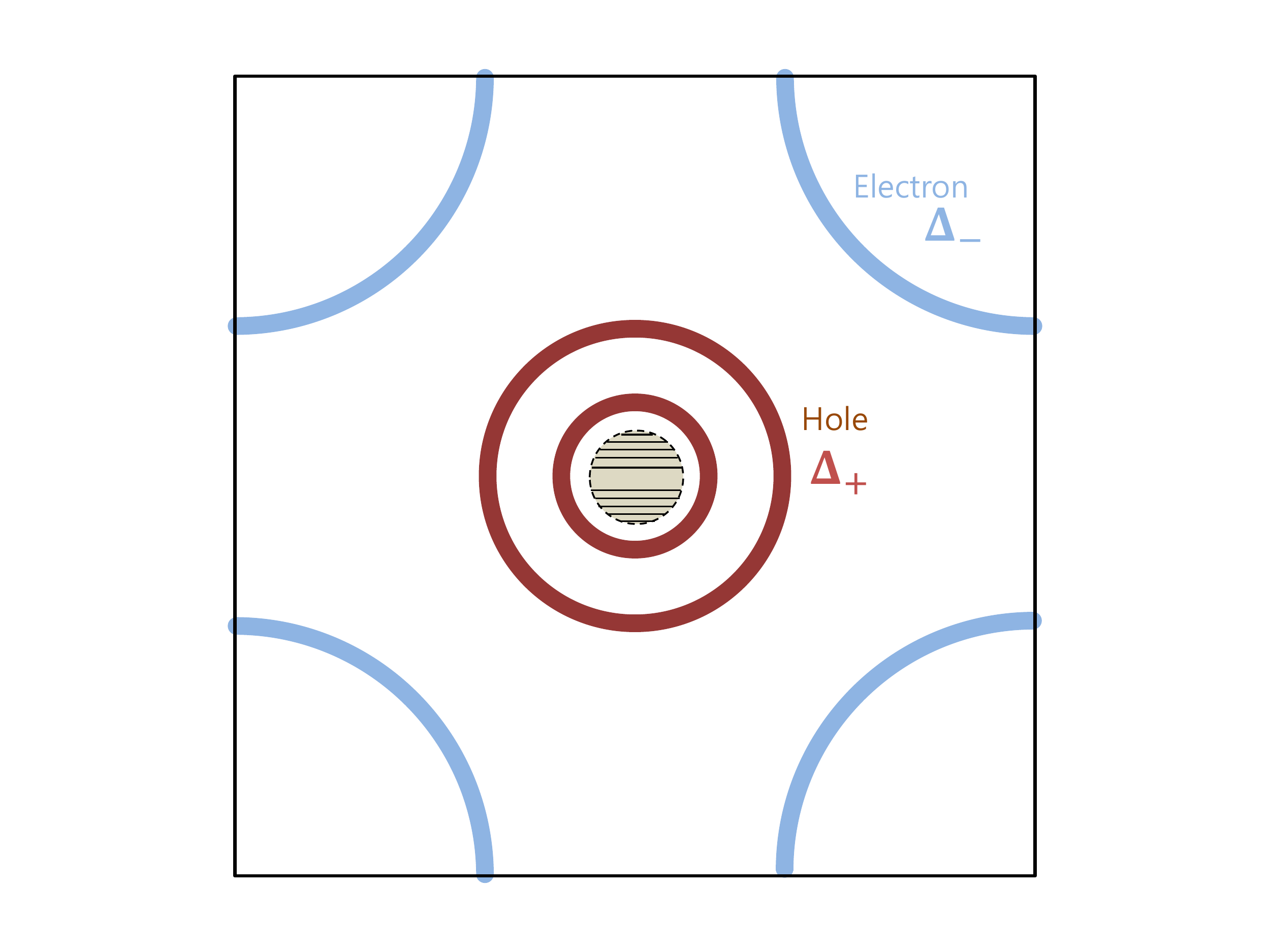}
\caption{Fermi surface of a typical iron based superconductor. Two hole pockets at the center
($k_F=0.1 \sim 0.3(\pi/a)$) and particle pockets at the corners have superconducting order parameters of opposite sign. There are five sub-bands in a nanowire with $w=50$nm and $k_F=0.04$. The 10 lines inside the dashed
circle correspond to 5 sub-bands in the nanowire and their endpoints are
the Fermi points of these sub-bands.}
\label{fig:iron}
\end{figure}
For $a=4\text{\AA}$ and either $w=50$ or $100$nm with 5 sub-bands occupied,
there is a mismatch between the Fermi momentum in the nanowire,
which is about 0.02 to 0.04 $(\pi/a)$, and the Fermi momentum in the pnictide.
This reduces the induced paring potential, as we will see in detail below.
For simplicity, we neglect the momentum dependence of the
gap in the inner hole pocket, $\Delta_k=\Delta_0=12$meV, as observed
in Ref. \onlinecite{Ding08} and assume momentum-conserving tunneling,
$t^{\rm u}(k,q)$, as introduced in Section \ref{sec:tunneling}.
We see from Eq. (\ref{eqn:induced-pairing}) that
\begin{align}
{r_{{k_x},i,j}} = \int \frac{dq_y}{(2\pi)^2}\frac{\left|t\right|^2 {g_i}({q_y}){g_j^*}({q_y})
}{\xi_{{k_x},{q_y}}^2+\left|\Delta_0\right|^2},\\
{\Delta'_{{k_x},i,j}} = \int \frac{dq_y}{(2\pi)^2}\frac{\left|t\right|^2 {g_i}({q_y}){g_j^*}({q_y})
\,{\Delta_0}}{\xi_{{k_x},{q_y}}^2+\left|\Delta_0\right|^2}.
\end{align}
We find that ${r_{{k_x},i,j}}$ and ${\Delta'_{{k_x},i,j}}$ are essentially momentum independent and diagonal in sub-band indices $i,j$ (see Appendix A), so we can assume that
the induced gap behaves like an $s$-wave superconducting gap. Using Eq. (\ref{eq:bdg}) with $\Delta(x,x')=\Delta^{ind}_i$ and $\mu_0=\mu_i$, we get a quartic equation for $z$ for each sub-band:
\begin{equation}
\frac{1}{4}z^4+\left(\tilde{\mu_i}+\tilde{\alpha}^2\right)z^2
-2\lambda\tilde{\Delta}_i\tilde{\alpha}z+\tilde{\mu_i}^2-\tilde{V}_x^2+\tilde{\Delta}^2_i=0,
\end{equation}
where $\tilde{x}=\frac{m^\ast\alpha x}{\hbar^2}$, $\tilde{\mu_i}=\frac{\hbar^2\mu_i}{m^\ast\alpha^2}$, $\tilde{V}_x=\frac{\hbar^2 V_x}{m^\ast\alpha^2}$ and $\tilde{\Delta}_i=\frac{\hbar^2}{m^\ast a^2}{\Delta^{ind}_i}$ with $u_{\uparrow,\downarrow}\propto e^{z\tilde{x}}$.
In this case, the condition for the $i^{\rm th}$ sub-band to be in the
topological phase is given by\cite{Lutchyn10}
$\tilde{\mu_i}^2-\tilde{V}_x^2+\tilde{\Delta}^2_i<0$.
Of course, our main concern is that the highest occupied sub-band
(or an odd number of sub-bands) be in the topological phase.

We now compute the induced pairing potentials for the parameters
described above. We consider a $w=50$nm wire where $i=1,2,3,4$ sub-bands are occupied,
and only one chirality-split branch of the $i=5$ sub-band is occupied. We set $\mu_5=5$meV and $V_x=15$meV to ensure the topological phase only in the 5th sub-band. For $t=60$meV, we get $r_{i}\sim0.12$ which does not renormalize the effective action much. 
Then the diagonal element of induced gap in each sub-band is (in meV):
\begin{align}
\label{eqn:pnictide-induced-pairing}
{\Delta}^{\rm ind}_{1} &=1.37 \\
{\Delta}^{\rm ind}_{2} &=1.37 \\
{\Delta}^{\rm ind}_{3} &=1.37 \\
{\Delta}^{\rm ind}_{4} &=1.38 \\
{\Delta}^{\rm ind}_{5} &=1.38. 
\end{align}

The topological phase transition occurs at $V_x=\sqrt{\mu_5^2+|\Delta^{ind}_5|^2}$. Therefore, the magnetic field required to observe the topological phase transition is a few tesla if $g\sim50$.
Since the iron pnictide superconductors have large $H_{c2}\sim50$T, the induced gap will not be significantly suppressed by the applied magnetic field in the region of phase transition.

Although the induced pairing is large, as may be seen from
Eq. (\ref{eqn:pnictide-induced-pairing}), the resulting single-particle
gap, which is the most physically-relevant quantity,
is significantly smaller according to Eq. (\ref{eqn:qpgap}).
For ${V_x}=15$ meV, $\Delta^{\rm qp} = 0.16$ meV.

The other types of interfaces are less interesting for pnictide superconductors.
At a dirty interface, electrons from the nanowire will be able to tunnel to all of the
components of the Fermi surface, which will suppress the induced gap since the electron
pockets are expected to have superconducting gaps of the opposite sign. A step-edge interface
could, similarly, allow tunneling to the electron pocket, which will suppress the induced gap.


\section{Superconductors with $d_{x^2 - y^2}$ Pairing Symmetry}

\subsection{Clean, Uniform Interface between a Cuprate Superconductor
and a Semiconductor Nanowire}

In this section we consider superconductors with $d_{x^2 - y^2}$ pairing symmetry,
such as the hole-doped cuprates. Then
${\Delta_q} = \frac{\Delta_0}{2}\left(\cos{q_x}a - \cos{q_y}a\right)$.
First, let us assume that the interface between the SC and the NW is clean and flat and, therefore,
momentum conserving. The direction of the NW is assumed to be parallel to the $a$-axis of the SC.
For the simplest possible form, ${t_i}(k_x,\textbf{q})= t_i^{\text u}(k_x,\textbf{q})$, we see that
\begin{equation}
{r_{{k_x},i,j,m}} =
 \int \frac{dq_y}{(2\pi)^2}
\frac{\left|t\right|^2{g_i}(q_y){g_j^*}(q_y)}{\omega_m^2+\xi_{{k_x},{q_y}}^2+\left|\Delta_{{k_x},{q_y}}\right|^2}
\label{eqn:gapeq0}
\end{equation}
and
\begin{multline}
{\Delta'_{{k_x},i,j,m}} =
 \int \frac{dq_y}{(2\pi)^2} \left|t\right|^2{g_i}(q_y){g_j^*}(q_y) \,\times\\
\frac{ \frac{\Delta_0}{2}\left(\cos{k_x}a - \cos{q_y}a\right)}{\omega_m^2+\xi_{{k_x},{q_y}}^2+\left|\Delta_{{k_x},{q_y}}\right|^2}
\label{eqn:gapeq1}
\end{multline}
Since ${g_i}({q_y})$ is peaked at $\pm \pi i/w$,
the momenta $({k_x},{q_y})$ will be far from the Fermi surface of
the superconductor for the first few nanowire sub-bands.
Therefore, ${\Delta'_{{k_x},i,j,m}}$ will be strongly suppressed.
Focussing, again, on the case of four and a half occupied sub-bands with $\mu_5=1$meV and $V_x=2$meV in the
nanowire with $w=50$nm, we find $r_{k_x,i,j}$ has very weak dependence on $k_x$
and is almost diagonal in sub-band index. We obtain $r_i\sim0.47$ for $t=1$eV.
\begin{figure}[h]
\includegraphics[width=8cm]{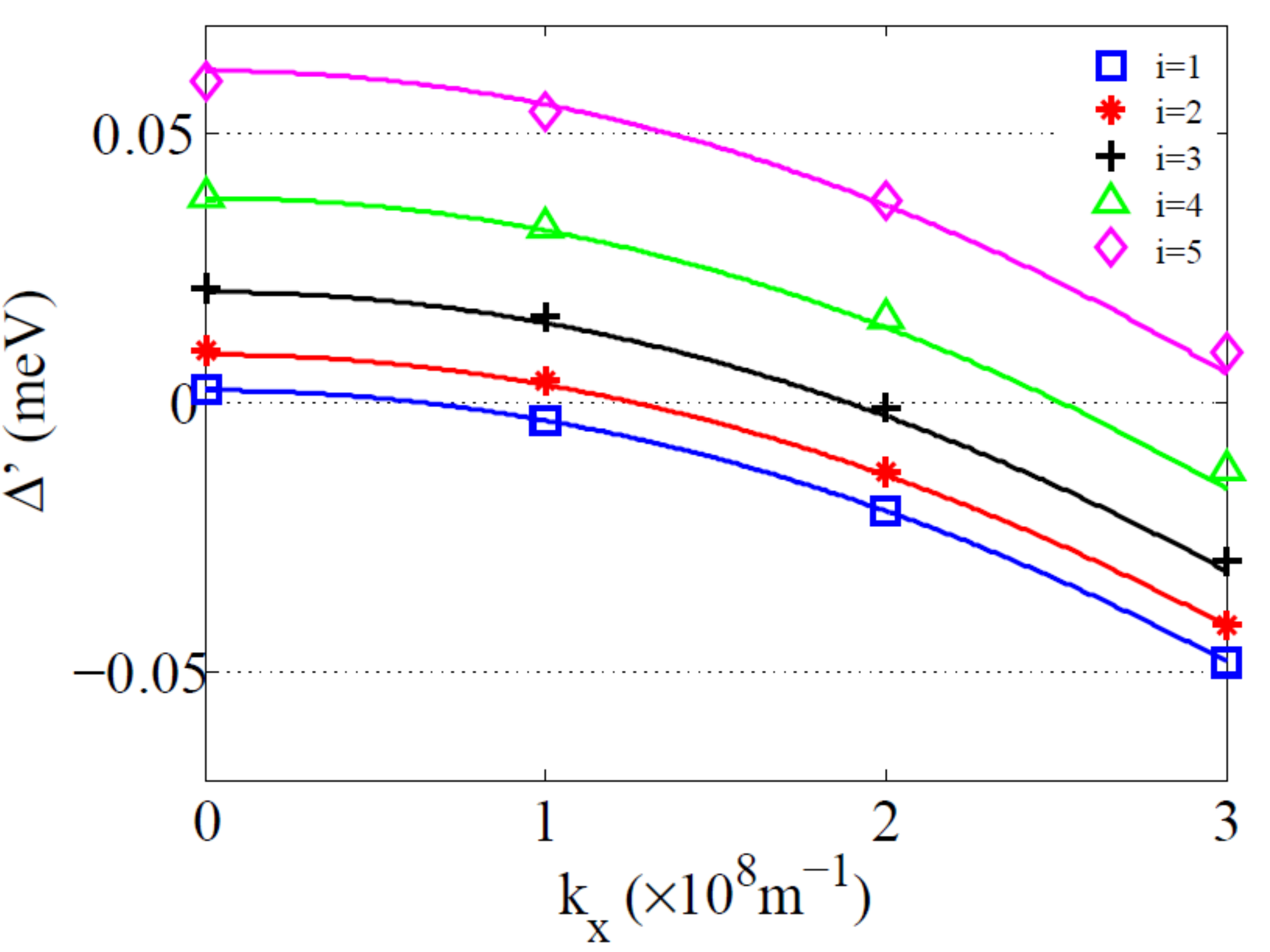}
\caption{Induced gap as a function of $k_x$ in each sub-band. Dots denote
the integration values. Plots of the $k_x$ dependence of a $d$-wave gap
(with $k_y$ fixed to the value expected for the corresponding sub-band)
are presented for guidance.}
\label{fig:gap-vs-k}
\end{figure}
We also find that the induced gap for each sub-band has the $k_x$ dependence
expected of a $d$-wave SC with $k_y$ fixed to the value expected for
each sub-band, as one can see in Fig. \ref{fig:gap-vs-k}.
The value of the induced gap at the Fermi momentum of each sub-band is (in meV)
\begin{align}
{\Delta}^{\rm ind}_{{k_F},1} &= -0.052\\
{\Delta}^{\rm ind}_{{k_F},2} &= -0.038\\
{\Delta}^{\rm ind}_{{k_F},3} &= -0.014\\
{\Delta}^{\rm ind}_{{k_F},4} &= 0.018\\
{\Delta}^{\rm ind}_{{k_F},5} &= 0.059.
\end{align}
For $V_x=2$ meV, this gives $\Delta^{\rm qp}=0.02$.

\begin{figure}[h]
\includegraphics[width=9cm]{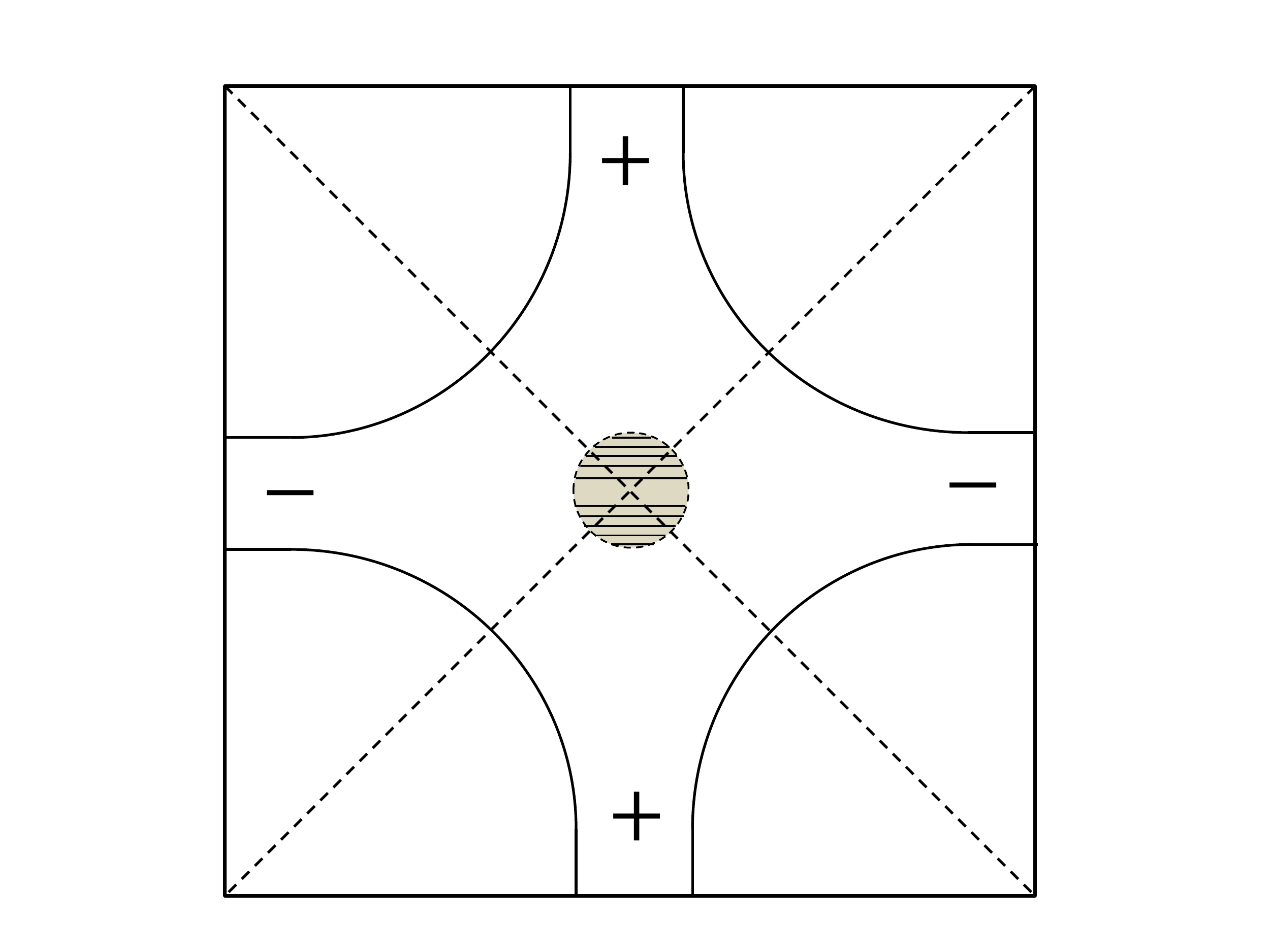}
\caption{Fermi surface of a cuprate superconductor. The 10 lines inside the dashed
circle correspond to 5 sub-bands in the nanowire. The ends of the lines are the Fermi
points of the nanowire, which are located far from the Fermi surface of the SC,
as may be seen from the figure.}
\label{fig:cuprate}
\end{figure}

We now consider Majorana zero modes at the end of such a nanowire. If we make
the approximation $g({q_y})\approx \delta\left({q_y}-\mbox{$\frac{\pi a}{w}$}\right)
+\delta\left({q_y}+\mbox{$\frac{\pi a}{w}$}\right)$ following Eq. (\ref{eqn:induced-pairing2}),
we can take $\Delta(x,x')\approx
{W_0}\left({a^2}\partial_x^2 - \left(\mbox{$\frac{\pi a}{w}$}\right)^2\right)$.
The BdG equation can then be written as in Eq. (\ref{eq:bdg}) with:
\begin{equation}
V_{\pm}\equiv
V_x\pm\lambda{W_0}\left({a^2}\partial_x^2 + \left(\mbox{$\frac{\pi a}{w}$}\right)^2\right)
\pm\alpha\partial_x
\end{equation}

The above BdG equation can be written as a quartic equation for $z$ with real coefficients.
\begin{multline}
\left(\frac{1}{4}+\tilde{\Delta}^2\right)z^4+2\tilde{\Delta}\lambda z^3+\left(1+\tilde{\mu}+\frac{2\pi^2\tilde{\Delta}^2}{\tilde{w}^2}\right)z^2\\
+\frac{2\pi^2\lambda\tilde{\Delta}}{\tilde{w}^2}z+\tilde{\mu}^2-\tilde{V}_x^2+\frac{\tilde{\Delta}^2\pi^4}{\tilde{w}^4}=0,
\end{multline}
where $\tilde{x}=\frac{m^\ast\alpha x}{\hbar^2}$, $\tilde{\mu}=\frac{\hbar^2\mu_0}{m^\ast\alpha^2}$, $\tilde{V}_x=\frac{\hbar^2 V_x}{m^\ast\alpha^2}$, $\tilde{\Delta}=\frac{m^\ast a^2}{\hbar^2}{W_0}$ and $\tilde{w}=\frac{m^\ast\alpha}{\hbar^2}w$. Note that when $z_{i}$s are the roots for
$\lambda=1$ channel, $-z_i$s are the solutions for $\lambda=-1$ channel. Since the coefficients are real, if $z_i$ is solution, $z_i^\ast$ is also a solution for same channel.\\
\begin{enumerate}

\item When $\tilde{\mu}^2-\tilde{V}_x^2+\frac{\tilde{\Delta}^2\pi^4}{\tilde{w}^4}<0$, there is at least one negative real root and one positive real root. Also, the product of the four roots $z_i$ is
$\frac{4(\tilde{\mu}^2-\tilde{V}_x^2+\frac{\tilde{\Delta}^2\pi^4}{\tilde{w}^4})}{1+4\tilde{\Delta}^2}<0$. When all roots are real, we have three positive roots for either $\lambda=1$ or $\lambda=-1$. When two of the roots are complex, then the four roots can be written $z_1>0$, $z_2<0$, $z_3=a+bi$ and $z_4=a-bi$, and we again have three roots with positive real part for either $\lambda=1$ or
$\lambda=-1$. Therefore, we have a MZM in this case.

\item When $\tilde{\mu}^2-\tilde{V}_x^2+\frac{\tilde{\Delta}^2\pi^4}{\tilde{w}^4}>0$ and all four roots are real, there are two different cases. When two of them are positive and two of them are negative, we do not have a localized solution for zero energy. When all four roots have same sign (which is positive for either $\lambda=1$ or $\lambda=-1$), we have two MZMs at the end of nanowire. However these two localized states are at the same end, and they will split into two states with $E>0$ and $E<0$ by interaction.

\item When $\tilde{\mu}^2-\tilde{V}_x^2+\frac{\tilde{\Delta}^2\pi^4}{\tilde{w}^4}>0$, and two roots are complex, the other two roots, if they are real, will have same sign since $\prod_{i=1}^4z_i>0$. If the other two roots are also complex, those two also have the same real part. Then it is similar to case 2.
 
\end{enumerate}

We do not consider the situation in which the above equation has a double root or two purely imaginary solutions in case 1 since those cases are sets of measure zero in parameter space. Therefore, we can conclude that there is a single Majorana bound state with zero energy when
$\tilde{\mu}^2-\tilde{V}_x^2+\frac{\tilde{\Delta}^2\pi^4}{\tilde{w}^4}<0$. Indeed, we can extend this analysis to the $n$th sub-band where $k_y=n\pi/w$,
and the condition for topological phase is given as
$\tilde{\mu}_n^2-\tilde{V}_x^2+\frac{\tilde{\Delta}^2n^4\pi^4}{\tilde{w}^4}<0$.

\subsection{Tunneling Through Cu $4s$ Orbitals}
We also consider the case in which
electrons tunnel from the nanowire only into Cu $4s$ orbitals.
These orbitals hybridize with neighboring Cu 3$d_{x^2 - y^2}$
orbitals to form a half-filled band at the Fermi energy.
Then $t_i(k_x,\textbf{q}) = t_i^{\text n}(k_x,\textbf{q})$.
We now have the intriguing form
\begin{multline}
{\Delta'_{k,i,j,m}} = \int \frac{dq_y}{(2\pi)^2}\left|t\right|^2 {g_i}(q_y){g_j^*}(q_y)\,\times\\
\times\frac{ \frac{\Delta_0}{2}\left(\cos{k_x}a - \cos{q_y}a\right)^3}{\omega_m^2+\xi_{{k_x},{q_y}}^2+
\left|\Delta_{{k_x},{q_y}}\right|^2}
\label{eqn:gapeq2}
\end{multline}
Although this expression is more sharply peaked at the anti-nodes,
where the gap is large, it suffers from the same limitation as the previous one,
namely that it is strongly suppressed by the Fermi momentum mismatch between
the nanowire and the superconductor. This is exacerbated by the fact
that nanowire electrons have momenta that are close to $k=(0,0)$ (on the scale
of the cuprate's Fermi momentum), and both $t(\textbf{k},\textbf{q})$ and $\Delta_k$ are very small there.
So, although this type of tunneling manages to avoid the nodes, it does not really
allow the nanowire electrons to tunnel to the superconductor's anti-nodes.
Consequently, the induced gaps are very small, as we show in more detail in
Appendix \ref{sec:Cu4s-numbers}.

\subsection{Dirty or Rough Interface}
We now consider the case of a dirty or rough interface. In the extreme case
introduced in Section \ref{sec:tunneling},  Eq. (\ref{eqn:induced-pairing}) becomes
\[
{\Delta'_{k,m}} = \int \frac{{d^2}\textbf{q}}{(2\pi)^2}\frac{\left|\lambda(\textbf{q})\right|^2 \, \Delta_q}{\omega_m^2+\xi_q^2+\left|\Delta_q\right|^2}
\]
The right-hand-side is independent of $k$. Moreover, since
$\Delta_q$ is odd under rotation by $\pi/2$ while the rest of the integrand is even,
the right-hand-side vanishes after integration, and there will be no induced gap.


\subsection{Nanowire - Step Edge Interface}\label{sec:analyzestepedge}
We now analyze the situation in which the nanowire is on top
of a step edge surface of a cuprate, as shown in Figure \ref{fig:step-edge}.
For simplicity, we assume that
the terraces are evenly spaced so that the terrace edges
are at $x_n = nl$. In the clean limit, the tunneling amplitude is dominant
at the terrace edges, $x_n$. We assume the tunneling matrix element
${t_i}(k_x,\textbf{q})={t_i^{\text s}}(k_x,\textbf{q})$ discussed in Sec II.
For small angle $\theta\sim10^{\circ}$, we can approximate $\cos\theta\sim1$, and
\begin{multline}
\Delta'_{k_x,k_x',i,j,m} = \sum_{n_1,n_2} \int \frac{d^2\textbf{q}}{(2\pi)^2}\left|t\right|^2\delta(k_x-q_x+n_1Q)\\
\times\delta(k_x'-q_x+n_2Q)\frac{g_i(q_y)g_j^\ast(q_y){\Delta_{q}} }{\omega_m^2+\xi_{q}^2+\left|\Delta_{q}\right|^2}
\end{multline}


At the densities under consideration in the NW, $k_F<<Q$, and we can ignore the contributions from $k_x=k_x'+(n_2-n_1)Q$ with $n_1\neq n_2$. By integrating over $k_x'$ as we did before, we get
\begin{multline}
\Delta'_{k_x,i,j,m} = \sum_{n} \int \frac{dq_y}{(2\pi)^2}\left|t\right|^2g_i(q_y)g_j^\ast(q_y)\\
\times\frac{{\Delta_{k_x+nQ,q_y}} }{\omega_m^2+\xi_{k_x+nQ,q_y}^2+\left|\Delta_{k_x+nQ,q_y}\right|^2}
\end{multline}
and
\begin{multline}
r_{k_x,i,j,m} = \sum_{n} \int \frac{dq_y}{(2\pi)^2}\frac{\left|t\right|^2g_i(q_y)g_j^\ast(q_y) }{\omega_m^2+\xi_{k_x+nQ,q_y}^2+\left|\Delta_{k_x+nQ,q_y}\right|^2}
\end{multline}
with $k_x+nQ\in$ first B.Z.\\
Again, in the limit of small inter-band coupling, the induced gap for each sub-band is given by
\begin{equation}
\Delta^{ind}_{k_F,i}=\frac{{\Delta'_{k_F,i,0}}}{1+r_{k_F,i,0}}.
\end{equation}
It is now possible for $({k_x}+nQ,{q_y})$ to lie on the Fermi surface of
the cuprate superconductor, so the induced gap is no longer suppressed.
\begin{figure}[h]
\includegraphics[width=9cm]{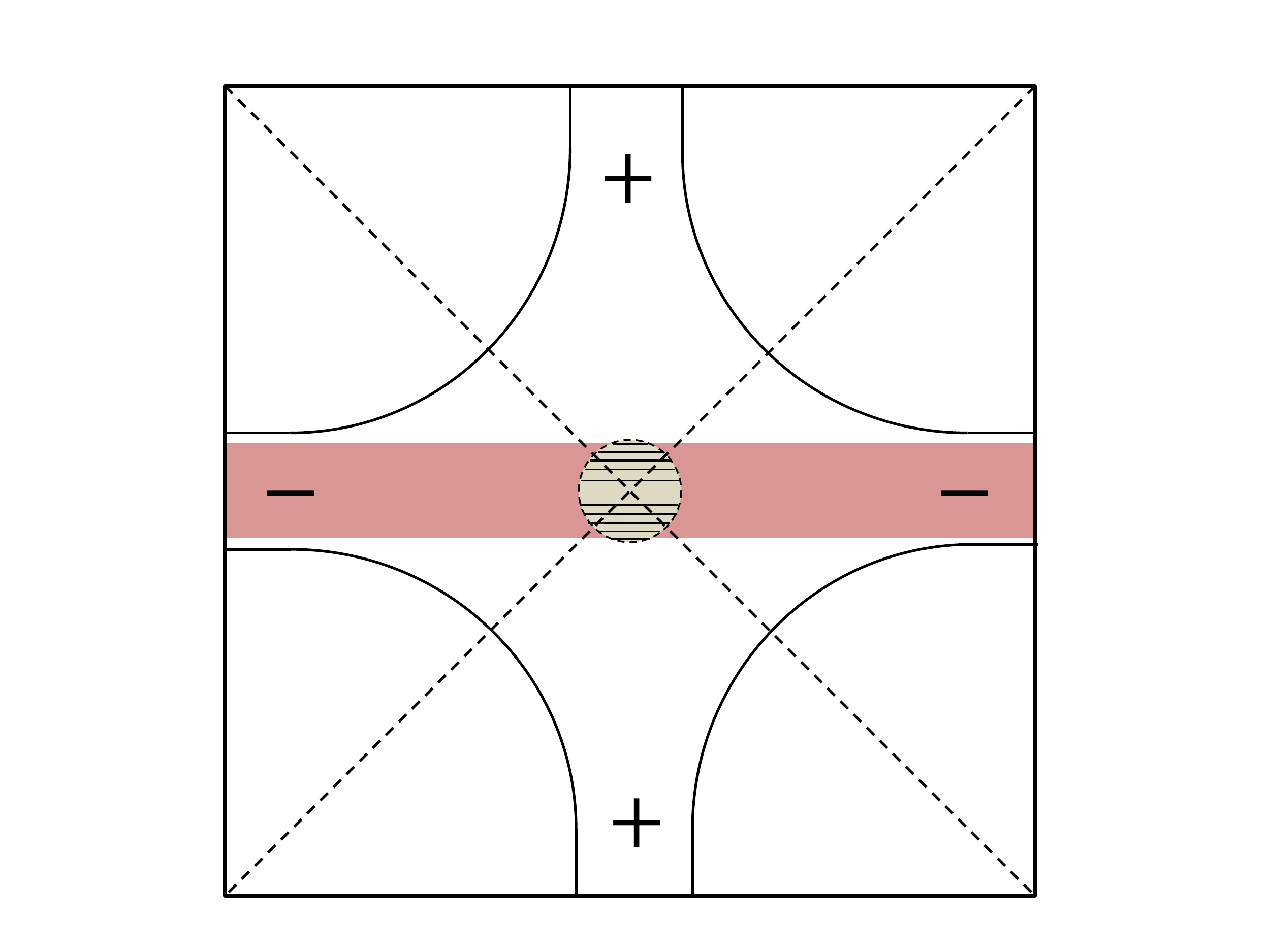}
\caption{Fermi surface of an interface between a semiconductor nanowire and a step-edge of a cuprate. Now induced superconducting gap gets contributions from the colored area.\label{fig:stepbz}}
\end{figure}


To estimate the size of the induced gap in each band, we take $\mu_5=10$meV and $V_x=30$meV for a nanowire of width $w=50$nm. Again we find $r_{k_x,i,j}$ is constant in $k_x$ and diagonal in $(i,j)$. Choosing $t=25$meV gives $r_i\sim0.11$ to $0.13$.
We find the induced gap for each sub-band has a very different $k_x$ dependence than
in the uniform tunneling case: the induced gap is only weakly-dependent on $k_x$
and, in particular, it does not change sign.
The induced gap in each sub-band at its Fermi momentum is
\begin{align}
{\Delta}^{\rm ind}_{{k_F},1} &= -3.04\\
{\Delta}^{\rm ind}_{{k_F},2} &= -3.09\\
{\Delta}^{\rm ind}_{{k_F},3} &= -3.15\\
{\Delta}^{\rm ind}_{{k_F},4} &= -3.20\\
{\Delta}^{\rm ind}_{{k_F},5} &= -2.64,
\end{align}
in meV. For $V_x=30$ meV, this gives $\Delta^{\rm qp}=0.2$.
In reality, tunneling will not be perfectly momentum-conserving modulo $Q$.
However, the basic result should still be valid: if momentum non-conservation is much
larger in one direction than the other then a large gap can be induced.

\begin{figure}[h]
\includegraphics[width=9cm]{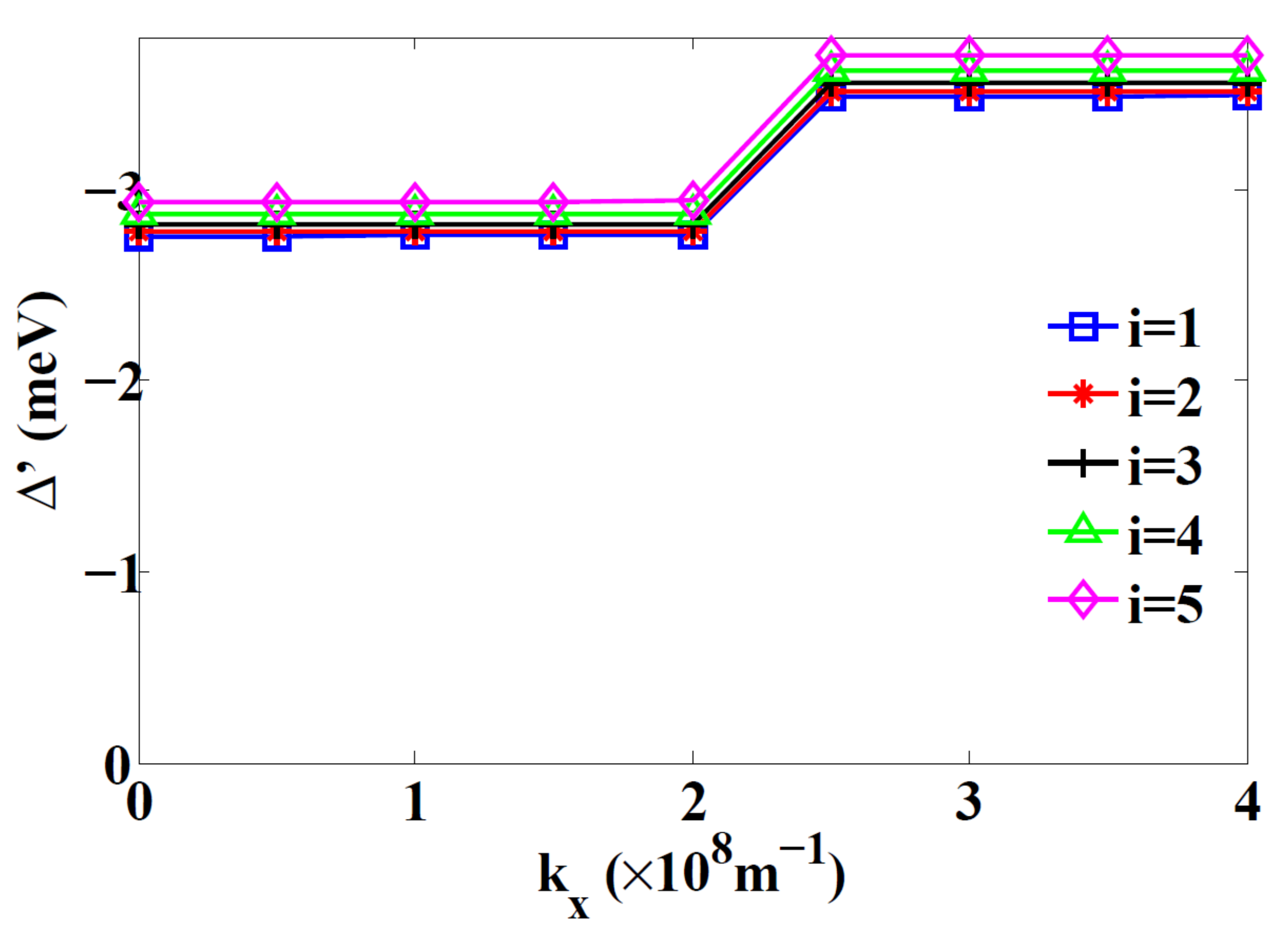}
\caption{Induced gap as a function of $k_x$ in each sub-band. Dots denote
the integration values. The lines are piecewise linear interpolations of the points.
The step occurs when one extra multiple of $Q$ contributes.
Note the contrast with Fig. \ref{fig:gap-vs-k}.}
\label{fig:gap-vs-kx-stepedge}
\end{figure}

We now consider Majorana zero modes at the end of the wire. As discussed in
Appendix \ref{sec:step-edge-numbers}, we find
the gap to vary very slowly with $k_x$ since the biggest contribution comes from $k_x+jQ\sim\pi/a$ where $\Delta_k$ and $\xi_k$ does not change rapidly with $k_x$.
Then, we are effectively in the case of an $s$-wave superconductor as we were in Sec III, and the characteristic equation for z
takes the form:
\begin{equation}
\frac{1}{4}z^4+\left(\tilde{\mu}_i+\tilde{\alpha}^2\right)z^2
-2\lambda\tilde{\Delta}_i\tilde{\alpha}z+\tilde{\mu}_i^2-\tilde{V}_x^2+\tilde{\Delta}_i^2=0.
\end{equation}
By essentially the same analysis as above, there are three solutions with
positive real part so long as $\tilde{\mu}_i^2-\tilde{V}_x^2+\tilde{\Delta}_i^2<0$.


\section{Mixing Between the Majorana Zero Mode and Gapless Bulk Excitations
in a $d_{x^2 - y^2}$-wave SC}
\label{sec:MZM-decay}

High-Tc superconductors with d-wave pairing symmetry have four gapless nodes in the
two-dimensional Brillouin zone. Thus far, we have ignored these low-energy
excitations because the tunneling matrix elements $t({k_x},\textbf{q})$ that we used
did not couple electrons in the nanowire to these nodal excitations so long as
the wire was infinitely-long. In a finite-length wire, however, momentum
along the wire is not conserved, and there will be some coupling
between low-energy electrons in the wire -- especially the zero modes -- and
nodal excitations in the superconductor. In addition, impurities could
scatter nanowire electrons to the nodal points. In this section we add a tunneling term to the action coupling the Majorana zero mode to the superconductor. We then calculate the self-energy of the Majorana mode perturbatively to analyze whether it will survive or decay into the bulk.

The tunneling action coupling the $j^{\rm th}$ sub-band Majorana zero mode localized at the end of the wire at $x=0$ to the fermionic excitations in the superconductor is
\begin{align}
S^{\gamma_j}_{T} &=\sum_m\int_{-\infty}^0 dx \int_0^w dy \int d^2\textbf{r}' v(\textbf{r},\textbf{r}')\gamma_{m,j}(\textbf{r})\nonumber\\
&\times\left[c^\sigma_{-m}(\textbf{r}')-\bar{c}^\sigma_{-m}(\textbf{r}')\right]\nonumber\\
&= \int_\textbf{k} v_{\gamma_j}(\textbf{k})\,\gamma_j(\omega_m)\left[c^{\sigma}_{-m,-\textbf{k}}-\bar{c}^{\sigma}_{-m,-\textbf{k}}\right]
\end{align}
where $v(\textbf{r},\textbf{r}')$ is the tunneling amplitude from the Majorana zero mode
at $r$ to a state at $r'$ in the superconductor. We take $v(\textbf{r},\textbf{r}')$ to be real. On the superconductor operators we have explicitly written the spin superscript $\sigma$. In going from the first to the second equality, we have assumed a simplified form for the real Majorana zero mode:
\begin{equation}
\gamma_j(\omega_m,\textbf{r})=2\sqrt{\frac{z}{w}}\sin\left(\frac{\pi jy}{w}\right)\Theta(-x)e^{zx}\gamma_j(\omega_m)
\end{equation}
We use the Fourier convention for $v_{\gamma_j}(\textbf{k})$
\begin{eqnarray}
v_{\gamma_j}(\textbf{k})=2\sqrt{\frac{z}{w}}\int_{-\infty}^0 dx \int_0^w dy \int d^2\textbf{r}' \,v(\textbf{r},\textbf{r}')\\
\times \sin\left(\frac{\pi jy}{w}\right)e^{zx+i\textbf{k}\cdot\textbf{r}'}
\end{eqnarray}
Then the self energy  $\Sigma_{\gamma_j\gamma_j}(\omega_m) =\langle \gamma_j(\omega_m)\gamma_j(-\omega_m) \rangle$ is given by:
\begin{align}
\Sigma&_{\gamma_j\gamma_j}(\omega_m) = \int d^2\textbf{k}\,\left|v_{\gamma_j}(\textbf{k})\right|^2\,
\Bigl[\left\langle c^{\sigma}_{\omega,\textbf{k}}c^{\sigma'}_{-\omega,-\textbf{k}}\right\rangle \nonumber\\
&\quad +\left\langle \bar{c}^{\sigma}_{\omega,-\textbf{k}}\bar{c}^{\sigma'}_{-\omega,\textbf{k}}\right\rangle
-\left\langle \bar{c}^{\sigma}_{\omega,-\textbf{k}} c^{\sigma'}_{-\omega,\textbf{k}}\right\rangle
-\left\langle c^{\sigma}_{\omega,\textbf{k}} \bar{c}^{\sigma'}_{-\omega,-\textbf{k}}\right\rangle\Bigr]\nonumber \\
&=2\int d^2\textbf{k} \frac{{\left|v_{\gamma_j}(\textbf{k})\right|^2}(\Delta_k+i\omega_m)}{\omega_m ^2+\xi_k ^2+\Delta_k ^2}
\end{align}
For simplicity, we take $\xi_k$ to be of the form
$\xi_k = t_1(\cos{k_x}a+\cos{k_y}a)/2 + t_2\cos{k_x}a\cos{k_y}a$
and $\Delta_k=\Delta_0(\cos k_xa-\cos k_ya)$. We consider three types of tunneling corresponding to the three interfaces described in Sec.~\ref{sec:tunneling}, for which the tunneling matrix elements are given explicitly by
\begin{eqnarray}
v^u_{\gamma_j}(\textbf{k})&=&\frac{t\sqrt{z}g_j(-k_y)}{z+ik_x}\\
v^n_{\gamma_j}(\textbf{k})&=&\frac{t\sqrt{z}g_j(-k_y)}{z+ik_x}2(\cos k_xa -\cos k_ya)\\
v^s_{\gamma_j}(\textbf{k})&=&\frac{td\sqrt{z}g_j(-k_y)}{1-e^{-(z+ik_x\cos\theta)l}}
\end{eqnarray}
\subsection{Zero Temperature Self Energy}
We obtain the retarded self-energy by taking $i\omega_m\rightarrow \omega+i\eta$ and then use the identity $\lim_{\eta \rightarrow 0+} \frac{1}{(x+i\eta)}=P(\frac{1}{x})-i\pi\delta(x)$ to obtain
\begin{equation}
\text{Im}\Sigma^{\gamma\gamma}_{r}(\omega) = 2 \int_{\textbf{k}\in \textbf{k}_0} \left|v_{\gamma_j}(\textbf{k})\right|^2
(\omega+\Delta_k)\left|\nabla_k(\xi_k ^2+\Delta_k^2)\right|^{-1} 
\label{eqn:ImSE}
\end{equation}
where $\textbf{k}_0$ satisfies $\omega^2 = \xi_{k_0}^2+\Delta_{k_0}^2$, and
\begin{equation}
\text{Re}\Sigma^{\gamma\gamma}_{r}(\omega) = \lim_{\eta \to 0}2\int d^2\textbf{k}\,
\left|v_{\gamma_j}(\textbf{k})\right|^2\frac{(\Delta_k+\omega)(\xi_k ^2+\Delta_k ^2-\omega ^2)}{(\omega ^2-\xi_k ^2-\Delta_k ^2)^2+\eta^2}
\label{eqn:ReSE}
 \end{equation}
From here on, $\Delta_k$ disappears from the numerator because it is odd under exchange of $k_x$ and $k_y$, while all other terms are even. We now explicitly calculate the real and imaginary parts. 

\subsubsection{Imaginary Part of Self Energy}\label{sec:imSE}

For small $\omega$, the dominant contribution to Eq. (\ref{eqn:ImSE}) comes from momenta near the nodes, which we denote by overbars: $(\pm \bar{k}_x,\pm \bar{k}_y)$. We expand the momenta around the nodal point $(\bar{k}_x,\bar{k}_y)$
as $({k_x},{k_y})=(\bar{k}_x + p + q,\bar{k}_y + p - q)$ and expand similarly
around the three other nodal points. Expanding $\xi_k$ and $\Delta_k$ about the nodal points yields $\xi_k=c_1p$ and $\Delta_k=c_2q$, where $c_1=-t_1 a\sin\bar{k_x}a-2t_2a\sin\bar{k}_x a\cos\bar{k}_ya$ and $c_2=2\Delta_0a\sin\bar{k}_x a$. Now the condition $\textbf{k}\in \textbf{k}_0$ is given by $\omega^2 = c_1 p^2 + c_2 q^2$.
We also linearize the tunneling strengths:
\begin{eqnarray}
v^u_{\gamma_j}(\textbf{k})&=&\frac{t\sqrt{z}g_j(-\bar{k}_y)}{z+i\bar{k}_x}=v^u\\
v^n_{\gamma_j}(\textbf{k})&=&-\frac{t\sqrt{z}g_j(-\bar{k}_y)}{z+i\bar{k}_x}\, qa \,\sin\bar{k}_xa=v^nqa\\
v^s_{\gamma_j}(\textbf{k})&=&\frac{td\sqrt{z}g_j(-\bar{k}_y)}{1-e^{-(z+i\bar{k}_x\cos\theta)l}}=v^s
\end{eqnarray}
At two of the nodal points, $q$ is replaced by $p$ in $v^n_{\gamma_j}$.
Note that
\begin{equation}
\left|v^{u,n}\right|^2 \propto a^4 z/w \, ,
\hskip 1 cm
\left|v^{s}\right|^2 \propto a^2d^2z/w
\end{equation}
are small because $100 < w/a < 250$. In addition,
$d\sim a$ and $1/z \gg a$, which further suppresses these tunneling
parameters.
The cases of uniform tunneling and a step-edge interface can be handled together:
\begin{eqnarray}
\text{Im}\Sigma^{\gamma\gamma}_r(\omega)&=& \frac{\left|v^{u,s}\right|^2}{\sqrt{2}}\omega \int_{\textbf{k}\in \textbf{k}_0}\frac{1}{\sqrt{c_1^4 p^2+c_2^4 q^2}}
\label{eqn:w3}
\end{eqnarray} A simple change of variables yields the decay rate $\Gamma(\omega)$ to leading order:
\begin{equation}
\Gamma^{u,s}(\omega) \propto \left(\frac{s}{w}\right)\, \frac{\omega}{\Delta_0} \, \frac{\left|t\right|^2}{\alpha t_1+\beta t_2}
\end{equation}
where $\alpha$ and $\beta$ are dimensionless numbers, and $s=a^2z$ or $d^2z$ in the cases of uniform and step-edge tunneling, respectively.
In the case of tunneling through the Cu 4s orbital, we have:
\begin{eqnarray}
\text{Im}\Sigma^{\gamma\gamma}_r(\omega)&=& \frac{\left|{v^n}\right|^2a^2}{\sqrt{2}}\omega \int_{\textbf{k}\in \textbf{k}_0}\frac{ q^2}{\sqrt{c_1^4 p^2+c_2^4 q^2}}
\label{eqn:w3}
\end{eqnarray}
so that
\begin{equation}
\Gamma^{n}(\omega) \propto \left(\frac{a^2z}{w}\right)\, \frac{\omega^3}{\Delta^3_0}\, \frac{\left|t\right|^2}{\alpha t_1+\beta t_2}
\end{equation}
In all cases, the decay rate is suppressed by a small coefficient. The decay rate for tunneling through the Cu 4s orbital is further suppressed at low energies by its cubic dependence on $\omega$.


\subsubsection{Real Part of Self Energy}
The real part of the self-energy follows similarly:
\begin{eqnarray}
Re\Sigma^{\gamma\gamma}_{r}(\omega)&= &\lim_{\eta \to 0} 2\int dpdq \left|v^{u,s}\right|^2\frac{\omega (c_1^2p^2 + c_2^2 q^2 -\omega^2)}{(c_1^2 p^2+c_2^2 q^2 -\omega^2)^2+\eta^2} \nonumber \\
&=&-\frac{2\left|v^{u,s}\right|^2\omega}{|c_1c_2|}\ln(\omega/\sqrt{\Lambda^2-\omega^2})
\end{eqnarray}
where $\Lambda$ is a high energy cut-off.
Hence, in the small $\omega$ limit,
the real part of the self-energy is logarithmically-divergent. Therefore,
the would-be pole at zero energy in the zero-mode Green function
has zero weight. This is a sign that the zero mode does not survive the
coupling to nodal quasiparticles. Eventually, the zero mode will leak into the
bulk of the $d_{x^2 - y^2}$ superconductor. However, the divergence is only
logarithmic because there is little phase space at the nodes, so this leakage
occurs very slowly. Moreover, the coefficient in front of this divergence is
small for the reasons noted above.

For tunneling through the Cu 4s orbital:
\begin{eqnarray}
Re\Sigma^{\gamma\gamma}_{r}(\omega)&= &\lim_{\eta \to 0} \int dpdq\frac{2\left|v^n\right|^2a^2}{|c_2^2|}\frac{\omega c_2^2q^2(c_1^2 p^2+c_2^2q^2 -\omega^2)}{(c_1^2p^2+c_2^2q^2 -\omega^2)^2+\eta^2} \nonumber \\
&=&\frac{2\left|v^n\right|^2a^2\omega}{|c_1c_2^3|}\left(\frac{\Lambda^2}{2}-\omega^2\ln(\omega/\sqrt{\Lambda^2-\omega^2})\right)
\end{eqnarray}
Hence, in the small $\omega$ limit, the real part of the self-energy goes like $\omega$. There is no divergence because tunneling
gets weaker as the nodes are approached. Therefore, there is a pole at zero energy
and the zero mode survives.


\subsection{Lifetime at Non-Zero Temperature}
At non-zero temperature, the preceding calculation is modified in two ways.
First, the Matsubara frequencies must take values $\omega_m = (2m+1)\pi/\beta$.
However, so long as we neglect electron-electron interactions in the
nanowire, there is no sum over Matsubara frequencies; the Matsubara
frequency of the zero mode is precisely the same as that of electrons or holes
in the superconductor. Therefore, this manifestation of non-zero temperature makes no difference.
The second way in which non-zero temperature can enter is through the
Green function of nodal fermions in the high-$T_c$ superconductor.
At non-zero temperature, these fermions have a lifetime proportional to $T$
(see, for instance Ref. \onlinecite{Valla99}).
Therefore, the superconductor's Green functions are modified according to
$\omega \rightarrow \omega - \Sigma_{SC}(\omega,T)$ with
$\text{Im} \Sigma_{SC}(\omega,T) \approx T$ for $\omega<T$.
Consequently, we now find that, for $\omega<T$:
\begin{eqnarray}
\Gamma^{u,s}(T) &\sim& T\ln T \nonumber\\
\Gamma^n(T) &\sim& T  \nonumber
\end{eqnarray}
The coefficients depend on the tunneling strengths $v$ in the same way as at zero temperature, so the decay rates will again be suppressed by numerical factors.

\section{Discussion}

Higher-$T_c$ superconductors are not a panacea for enhancing
the stability of Majorana zero modes.  Part of the beauty of
nanowire-superconductor proposals for Majorana zero modes
is that they are relatively insensitive to the details of the superconductor.
However, since the cuprates and pnictides (and, perhaps, all
higher-$T_c$ superconductors) have gap functions that
change sign in the Brillouin zone, their details necessarily matter.
Indeed one might, initially, expect that, as a result, it is impossible to
use them to induce robust topological superconductivity in a nanowire.
However, we have seen that it is possible, with the right type of interface, to do so.
In fact, for a nanowire in contact with a cuprate superconductor at
a step-edge surface, a pairing gap of $\approx 3$ meV can be induced,
and it is conceivable that a pairing gap as large as $15$ meV
could be induced with sufficiently strong tunneling at the interface.
Even in the case of a pnictide superconductor, a gap of $\approx 1$ meV
seems achievable.

In the case of the cuprates, there is a second difficulty: the presence of
nodal fermions. Naively, these should immediately wipe out the possibility
of topological superconductivity and Majorana zero modes. However,
the nanowire's electrons couple weakly to the nodal points, so the decay of
the zero modes may be slow. They will always decay into the bulk
(except in the special case in which tunneling occurs solely through
the Cu 4s orbital), but for $w/a$ sufficiently large, this may not be a quantitatively
larger effect than the decay into impurity states in an $s$-wave superconductor.

The primary focus in this work has been to increase the energy and temperature
scale associated with superconductivity. This could help make Majorana zero modes
more robust in smaller systems or higher temperatures. However, the spin-orbit
coupling strength is equally, if not more, important for determining the energy 
and length scales that protects the zero modes. We have seen that
$\Delta^{\rm qp}$ is consistently an order of magnitude smaller than
$\Delta^{\rm ind}$; this occurs because the spin-orbit energy is an order
of magnitude smaller than the Zeeman energy which, in turn, must be large
because the induced superconducting gap is large.
Therefore, it would be interesting to see
how the choice of interface geometry -- for instance, the terrace structure
that helps a $d_{x^2 - y^2}$-superconductor induce superconductivity
in a nanowire -- can help enhance spin-orbit coupling in the nanowire.

\acknowledgements
We would like to thank R. Lutchyn, E. Plamadeala and G. Y. Cho for discussions
and V. Galitski for sharing his and his collaborators' manuscript prior to publication.
This work has been partially supported by the DARPA QuEST program
and the AFOSR under grant FA9550-10-1-0524.

\appendix

\section{Numerical Values of the Gap and Renormalized Dispersion}

We present numerical values for the matrices $r$, $\epsilon'$ and $\Delta'$ defined in eqns~\ref{eqn:wave-renorm}-\ref{eqn:energy-renorm}, evaluated at the Fermi momenta of the highest filled band in several different cases. From now on, the units are $(t/1 eV)^2$ for $r$ and $t^2/(1 eV)$
for $\epsilon'$ and $\Delta'$. Values for the induced gap $\Delta^{ind}$
are presented in plots as a function of $t$.

\subsection{Pnictides: $s_{\pm}$ pairing}
\begin{figure}[h]
\includegraphics[width=8cm]{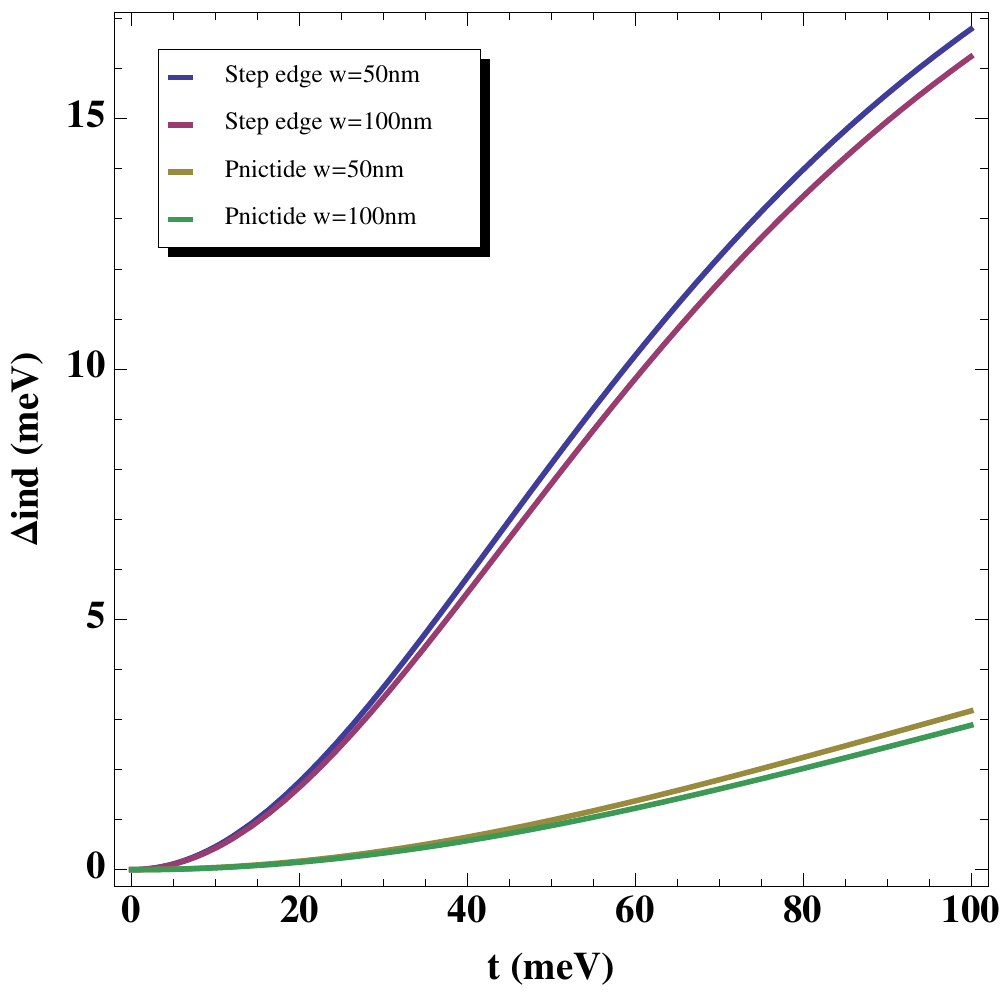}
\caption{Induced gap for a cuprate superconductor with step edge tunneling (top lines) and an
$s\pm$ superconductor (bottom two lines) as a function of tunneling strength. Within each pair the larger induced gap corresponds to the narrower wire.}
\label{fig:inducedgapPnicAndStep}
\end{figure}

We assume that the Fermi level lies between the chirality split bands of the fifth subband and take the representative values $\mu_5 = 5$meV and $V_x=15$meV. The Fermi momentum of the fifth band is $k_{F,5}=.086\text{nm}^{-1}$. For a wire of width $50$nm, we find the matrices at
$\omega_m=0$ to be:
\[r_{k_x=k_x'=k_{F,5}}=\left(
 \begin{tabular}{ccccc}
  30.6 & 0 & .036 & 0 & .062 \\
  0 & 31.2 & 0 & .098 & 0 \\
  .036 & 0 & 32.3 & 0 & .190 \\
  0 & .098 & 0 & 33.78 & 0 \\
  .062 & 0 & .190 & 0 & 35.9
 \end{tabular} \right)
\]
\[\epsilon '_{k_x=k_x'=k_{F,5}}=\left(
 \begin{tabular}{ccccc}
  1.52 & 0 & -.0003 & 0 & -.0005 \\
  0 & 1.53 & 0 & -.0008 & 0 \\
  -.0003 & 0 & 1.55 & 0 & -.0016 \\
  0 & -.0008 & 0 & 1.58 & 0 \\
  -.0005 & 0 & -.0016 & 0 & 1.62\\
 \end{tabular} \right)
\]
Since the pairing function is constant the matrix $\Delta'$ is proportional to the matrix
 $r$. The matrices evaluated at the Fermi momenta of the other bands yield similar values. Hence we conclude that
$r$ and $\epsilon'$ are diagonal in both $k_x$ and sub-band index (the latter follows
from the small values of the off-diagonal matrix elements above).
Consequently, we can find the induced gap for each band from Eq.~(\ref{eqn:inducedgap}) as a function of the tunneling strength $t$ (defined in Eq.~(\ref{eqn:uniformtunneling})), shown in Fig~\ref{fig:inducedgapPnicAndStep}. The figure also shows the induced gap for a wider wire of width $w=100$nm, for which the matrices $r$ and $\epsilon'$ are similar to the above matrices, but with smaller off-diagonal elements because $y$-momentum is more nearly a conserved quantity.
Since $r$, $\epsilon'$, and $\Delta'$ remain small even for $t$ as large as $100$meV,
it is possible that even such a large tunneling matrix element can be treated
as a small perturbation. However, even for more a more modest tunneling matrix element
such as $50$meV, the induced gap is approximately $1$ meV.

\subsection{Cuprates: uniform tunneling}
\begin{figure}[h]
\includegraphics[width=8cm]{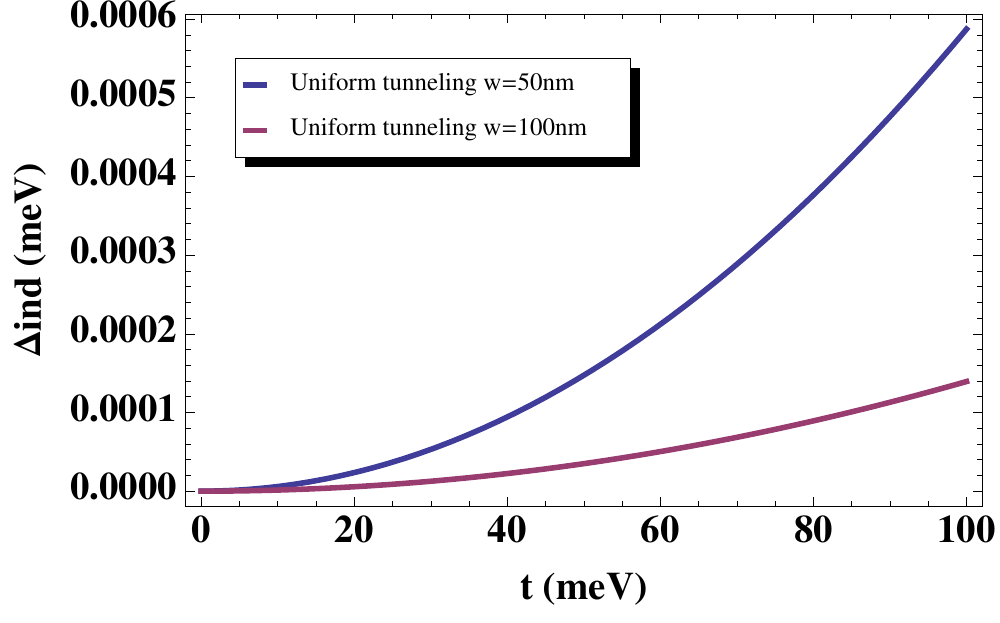}
\caption{Induced gap for a cuprate superconductor with uniform tunneling as a function of tunneling strength for wires of width $w=50$nm (top) and $w=100$nm (bottom)
\label{fig:inducedgapUniformCuprate}}
\end{figure}

We assume again that the Fermi level lies between the chirality split bands of the fifth subband and take the representative values $\mu_5=1$meV, $V_x=2$meV. The Fermi momentum of the fifth band is $k_{F,5}=.034\text{nm}^{-1}$. For a wire of width $w=50$nm we find the following
matrices at $\omega_m=0$:
\[r_{k_x=k_x'=k_{F,5}}=\left(
 \begin{tabular}{ccccc}
  .465 & 0 & .00002 & 0 & .00003 \\
  0 & .466 & 0 & .00004 & 0 \\
  .00002 & 0 & .467 & 0 & .00008 \\
  0 & .00004 & 0 & .469 & 0 \\
  .00003 & 0 & .00008 & 0 & .470
 \end{tabular} \right)
\]
\[\epsilon'_{k_x=k_x'=k_{F,5}}=\left(
 \begin{tabular}{ccccc}
  -192 & 0 & -.0004 & 0 & -.0007 \\
  0 & -193 & 0 & -.001 & 0 \\
  -.0004 & 0 & -193 & 0 & -.002 \\
  0 & -.001 & 0 & -193 & 0 \\
  -.0007 & 0 & -.002 & 0 & -193
 \end{tabular} \right) \times 10^{-3}
\]
\[\Delta'_{k_x=k_x'=k_{F,5}}=\left(
 \begin{tabular}{ccccc}
  17.1 & 0 & 4.69 & 0 & 7.83 \\
  0 & 88.1 & 0 & 12.5 & 0 \\
  4.69 & 0 & 207 & 0 & 23.5 \\
  0 & 12.5 & 0 & 374 & 0 \\
  7.83 & 0 & 23.5 & 0 & 590
 \end{tabular} \right) \times 10^{-7}
\]
\\
Similar matrices are obtained at the Fermi momenta of the other filled bands, although there is variation in $\Delta'$ with $k_x$ which can be positive or negative, but is always very small. The $r$ and $\epsilon'$ matrices are effectively diagonal, although $\Delta'$ is not. However, since the elements of $\Delta'$ are so much smaller than the energy splitting between bands, we can still treat the effective action in Eq.~(\ref{eqn:effective-action}) to be diagonal in bands to estimate the induced gap from Eq.~(\ref{eqn:inducedgap}). The induced gap is extremely small; we plot it as a function of the tunneling strength $t$ in Fig~\ref{fig:inducedgapUniformCuprate}, along with the induced gap for a wider wire with $w=100$nm.
\\
\\
\subsection{Cuprates: tunneling through Cu $4s$ orbitals}
\label{sec:Cu4s-numbers}
Using the same representative values $\mu_5=1$meV and $V_x=2$meV we can find the matrix elements for tunneling into Cu $4s$ orbitals described in Sec~\ref{sec:tunneling}. These elements are greatly suppressed by the factor $(\text{cos}(q_x a) - \text{cos}(q_y a))^2$ in Eq.~(\ref{eqn:neighbortunneling}) because of the momentum mismatch between the wire and the superconductor and are two or more orders of magnitude less than the uniform tunneling described in the previous section. Hence, we will not discuss this case in any more detail.
\\
\subsection{Cuprates: tunneling through the step edge interface}
\label{sec:step-edge-numbers}
Following Sec~\ref{sec:analyzestepedge}, we anticipate a sizable induced gap in this case because $k_x = q_x$ only modulo $Q$, so we choose a larger value $V_x = 30$meV to accommodate
this. We choose $\mu_5=10$meV. We again assume that the chemical potential lies between the chirality split levels of the fifth subband, which has Fermi momentum $k_{F5}=.12\text{nm}^{-1}$. We calculate matrix elements evaluated at this momentum and at $\omega_m=0$: 

\[r_{k_x=k_x'=k_{F,5}}=\left(
 \begin{tabular}{ccccc}
  169 & 0 & .050 & 0 & .085 \\
  0 & 169 & 0 & .126 & 0 \\
  .050 & 0 & 172 & 0 & .254 \\
  0 & .126 & 0 & 176 & 0 \\
  .085 & 0 & .254 & 0 & 180
 \end{tabular} \right)
\]
\[\epsilon'_{k_x=k_x'=k_{F,5}}=\left(
 \begin{tabular}{ccccc}
  -10.8 & 0 & .001 & 0 & .002 \\
  0 & -10.8 & 0 & .003 & 0 \\
  .001 & 0 & -10.9 & 0 & .005 \\
  0 & .003 & 0 & -11.0 & 0 \\
  .001 & 0 & .005 & 0 & -11.1
 \end{tabular} \right)
\]
\[\Delta'_{k_x=k_x'=k_{F,5}}=\left(
 \begin{tabular}{ccccc}
  -4.37 & 0 & -.0005 & 0 & -.0008 \\
  0 & -4.41 & 0 & -.001 & 0 \\
  -.0005 & 0 & -4.46 & 0 & -.002 \\
  0 & -.001 & 0 & -4.55 & 0 \\
  -.0008 & 0 & -.002 & 0 & -4.66
 \end{tabular} \right)
\]

We have summed over all $n$ such that $k_{F,5} + nQ$ lies within the first Brillouin zone of the superconductor ($|n| \leq 8$). Matrix elements evaluated at the Fermi momenta of other filled bands will be similar; there will be small differences when $k_{F,j}$ is such that other $n$ are allowed, but these will ultimately result in a very similar induced gap. We do not need to worry about
$k_x \neq k_x'$ because if $k_x'={k_x}+Q$, then only one of them can be near the Fermi
surface of the nanowire. Since the matrices are effectively diagonal, we can calculate the induced gap from Eq.~(\ref{eqn:inducedgap}), shown in Fig~\ref{fig:inducedgapPnicAndStep} as a function of the tunneling strength $t$, along with the induced gap for a wider wire with $w=100$nm.

\bibliography{Majorana}

\end{document}